\documentclass{LMCS}

\usepackage{graphicx}
\usepackage{amsmath}
\usepackage{amssymb}
\usepackage{latexsym}
\usepackage{amscd}
\usepackage{stmaryrd}
\usepackage{enumerate}
\usepackage{hyperref}

\newcommand{\ignore}[1]{}

\newcommand{\problem}[3]{%
  \vspace{2mm}
  \noindent
  \framebox{
    \begin{minipage}{0.94\hsize}
      \noindent{#1}

      \noindent{\bf Instance}

      \begin{itemize}
        #2
      \end{itemize}

      \noindent{\bf Task} {#3}
    \end{minipage}
  }
  \vspace{1mm}%

  \noindent%
}

\newcommand{\ctlstar}{$\mathit{CTL}^*$}
\newcommand{\hide}[1]{}
\newcommand{\tuple}[1]{\left(#1\right)}
\newcommand{\states}{S}
\newcommand{\fstates}{F}
\newcommand{\qstates}{Q}
\newcommand{\state}{s}
\newcommand{\initstate}{\state_{\it init}}
\newcommand{\unreachable}[1]{\widetilde{#1}}
\newcommand{\reachable}[1]{{#1}^{\bullet}}
\newcommand{\qualreach}{{\sc Qual\_Reach }}
\newcommand{\qualrepreach}{{\sc Qual\_Rep\_Reach }}
\newcommand{\aqrr}{{\sc Approx\_Quant\_Rep\_Reach }}

\newcommand{\ef}{\exists\Diamond}

\newcommand{\transition}[1]{\stackrel{#1}\longrightarrow}
\newcommand{\tstuple}{\tuple{\states,\transition{}}}
\newcommand{\probability}{P}
\newcommand{\probms}[1]{{\it Prob_{#1}}}
\newcommand{\tssystem}{{\mathcal T}}
\newcommand{\mchain}{{\mathcal M}}
\newcommand{\set}[1]{\left\{#1\right\}}
\newcommand{\setcomp}[2]{\left\{#1|\;#2\right\}}
\newcommand{\mctuple}{\tuple{\states,\probability}}
\newcommand{\pAth}{\pi}
\newcommand{\run}{\rho}
\newcommand{\pAthes}{\Pi}

\newcommand{\dist}[1]{{\it dist_{#1}}}

\newcommand{\post}{{\it Post}}
\newcommand{\eventually}{\Diamond}
\newcommand{\always}{\Box}
\newcommand{\until}{{\mathcal U}}

\newcommand{\rat}{\mathbb Q}
\newcommand{\nat}{\mathbb N}
\newcommand{\R}{{\rm I\hskip-0.21em R}}
\newcommand{\yes}{{\it Yes}}
\newcommand{\no}{{\it No}}
\newcommand{\before}{\;\underline{{\it Before}}\;}

\newcommand{\aqr}{{\sc Approx\_Quant\_Reach }}
\newcommand{\attractor}{A}
\newcommand{\eqr}{{\sc Exact\_Quant\_Reach }}
\newcommand{\error}{{\varepsilon}}
\newcommand{\probapprox}{{\theta}}
\newcommand{\probbound}{{\nu}}

\def\om{\omega}
\def\De{\Delta}
\def\Om{\Omega}

\newcommand{\vass}{{\mathcal V}}
\newcommand{\vassstate}{\lcsstate}
\newcommand{\vassstates}{\lcsstates}
\newcommand{\vasstransition}{\lcstransition}
\newcommand{\vasstransitions}{\lcstransitions}
\newcommand{\varstate}{{\tt v}}
\newcommand{\var}{{\tt x}}
\newcommand{\vassweight}{{\tt w}}
\newcommand{\vasstuple}{\tuple{\vassstates,\vars,\vasstransitions}}
\newcommand{\pvasstuple}{\tuple{\vassstates,\vars,\vasstransitions,\vassweight}}

\newcommand{\vars}{{\tt X}}
\newcommand{\pvass}{\vass}
\newcommand{\qvassstates}{\qlcsstates}

\newcommand{\ntm}{{\mathcal N}}
\newcommand{\ntmstate}{\lcsstate}
\newcommand{\ntmstates}{\lcsstates}
\newcommand{\ntmtransition}{{\tt t}}
\newcommand{\ntmtransitions}{{\tt T}}
\newcommand{\inputalphabet}{{\tt \Sigma}}
\newcommand{\tapealphabet}{{\tt \Gamma}}
\newcommand{\blank}{\sharp}
\newcommand{\noise}{\epsilon}
\newcommand{\tapes}{{\tt M}}

\newcommand{\ntmsymbol}{\sigma}
\newcommand{\move}{m}
\newcommand{\ntmweight}{{\tt w}}
\newcommand{\qntmstates}{{\tt Q}}
\newcommand{\ntmtuple}{\tuple{\ntmstates,\inputalphabet,\tapealphabet,\tapes,\ntmtransitions,\noise,\ntmweight}}
\newcommand{\graph}[1]{{\mathcal G}(#1)}

\newcommand{\lcs}{{\mathcal L}}
\newcommand{\plcs}{\lcs}
\newcommand{\lcsstates}{{\tt S}}
\newcommand{\qlcsstates}{{\tt Q}}
\newcommand{\lcsstate}{{\tt s}}
\newcommand{\msgs}{{\tt M}}
\newcommand{\msg}{{\tt m}}
\newcommand{\channels}{{\tt C}}
\newcommand{\channel}{{\tt c}}
\newcommand{\lcstransitions}{{\tt T}}
\newcommand{\lcstransition}{{\tt t}}
\newcommand{\op}{{\tt op}}
\newcommand{\nop}{{\tt nop}}

\newcommand{\transitionx}[1]{\overset{{#1}}{\longrightarrow}}
\newcommand{\chassignment}{{\tt x}}
\newcommand{\lcsweight}{{\tt w}}

\newcommand{\plcstuple}{\tuple{\lcsstates,\channels,\msgs,\lcstransitions,\lossp,\lcsweight}}
\newcommand{\lossp}{\lambda}

\newcommand{\enabled}{{\it enabled}}

\newlength{\thintextwidth}
\newcommand{\thintext}[1]{%
  \settowidth{\thintextwidth}{#1}%
  {#1}\hspace{-\thintextwidth}%
}
\newcommand{\ind}{${}$\hspace{0.5cm}}
\newcommand{\algline}[1]{\thintext{#1}\ind}
\newcommand{\algbeginspace}{\\[2mm]}

\newcounter{algboxcounter}
\newcommand{\algbox}[3]{%
  \refstepcounter{algboxcounter}
  \vspace{1mm}
  \noindent\framebox
  {
    \begin{minipage}{0.95\hsize}
      {\bf Algorithm \thealgboxcounter{} -- #1}\\
      #2
      \algbeginspace
      \noindent{\bf begin}\\
      #3\\
      \noindent{\bf end}
    \end{minipage}
  }
  \vspace{1mm}
}
\def\eqalign#1{\null\,\vcenter{\openup\jot\mathsurround=0 pt
  \ialign{\strut\hfil$\displaystyle{##}$&$\displaystyle{{}##}$\hfil
      \crcr#1\crcr}}\,}

\def\doi{3 (4:7) 2007}
\lmcsheading%
{\doi}
{1--32}
{}
{}
{Dec.~\phantom{0}6, 2006}
{Nov.~\phantom{0}8, 2007}
{}

\begin{document}
\title[Decisive Markov Chains]{Decisive Markov Chains\rsuper *}
\author[P.~A.~Abdulla]{Parosh Aziz Abdulla\rsuper a}
\address{{\lsuper a}Uppsala University, Department of Information Technology,
  Box 337, SE-751 05 Uppsala, Sweden}
\email{\{parosh,noomene.behenda\}@it.uu.se}
\author[N.~B.~Henda]{Noomene Ben Henda\rsuper a}
\address{\vskip-6 pt}
\author[R.~Mayr]{Richard Mayr\rsuper b}
\address{{\lsuper b}North Carolina State University,
Department of Computer Science,
Raleigh, NC 27695-8206, USA}
\email{mayr@csc.ncsu.edu}

\titlecomment{{\lsuper *}An extended abstract of an earlier version of this paper
has appeared in the proceedings of LICS 2005 \cite{ABM:LICS05}.}

\begin{abstract}
\noindent
We consider qualitative and quantitative verification problems 
for infinite-state Markov chains.
We call a Markov chain {\em decisive} w.r.t.\ a given set of
target states $\fstates$ if it almost certainly eventually reaches either 
$\fstates$ or a state from which $\fstates$ can no longer be reached.
While all finite Markov chains are trivially decisive 
(for every set $\fstates$), 
this also holds for many classes of infinite Markov chains.

Infinite Markov chains which contain a finite attractor
are decisive w.r.t.\ every set $\fstates$. In particular, 
all Markov chains induced by probabilistic lossy channel systems (PLCS)
contain a finite attractor and are thus decisive.
Furthermore, all globally coarse Markov chains are decisive.
The class of globally coarse Markov chains includes, e.g., those
induced by probabilistic vector addition systems (PVASS) 
with upward-closed sets $\fstates$, and all Markov 
chains induced by probabilistic noisy Turing machines (PNTM)
(a generalization of the noisy Turing machines (NTM) of
Asarin and Collins).

We consider both safety and liveness problems for decisive Markov chains.
Safety: What is the probability that a given set of states $\fstates$ is
eventually reached. Liveness: What is the probability that a given set of
states is reached {\em infinitely often}. There are three variants of these
questions. (1) The qualitative problem, i.e., deciding
if the probability is one (or zero); (2) the approximate quantitative problem,
i.e., computing the probability up-to arbitrary precision; 
(3) the exact quantitative problem, i.e., computing probabilities exactly.

1. We express the qualitative problem in abstract terms for 
decisive Markov chains, and show an almost complete picture of 
its decidability for PLCS, PVASS and PNTM.

2. We also show that the path enumeration algorithm of Iyer and Narasimha
terminates for decisive Markov chains and can thus be used to
solve the approximate quantitative safety problem.
A modified variant of this algorithm can be used to
solve the approximate quantitative liveness problem.

3. Finally, we show that the exact probability of (repeatedly) reaching $\fstates$
cannot be effectively expressed (in a uniform way) in Tarski-algebra
for either PLCS, PVASS or (P)NTM (unlike for probabilistic pushdown automata).
\end{abstract}

\maketitle

\section{Introduction}
\label{introduction:section}

\paragraph{\bf Verification of infinite systems.}
The aim of model checking is to decide algorithmically whether
a transition system satisfies a specification.
Specifications which are formulated as
{\it reachability}
or {\it repeated reachability} of a given set of target states are
of particular interest since they allow to analyze
{\it safety} and {\it progress} properties respectively.
In particular, model checking problems w.r.t. $\omega$-regular 
specifications are reducible to the repeated reachability problem.

A main challenge has been to
extend the  applicability of model checking 
to systems with infinite state spaces.
Algorithms have been developed for numerous
models such as timed automata, Petri nets, pushdown systems, 
lossy channel systems, parameterized systems, etc.

\paragraph{\bf Probabilistic systems.}
In a parallel development, methods have been
designed for the analysis of models with stochastic behaviors 
(e.g.\ \cite{
LeSh:chance,
HaSh:probabilistic,
Vardi:probabilistic,
CY:FOCS88,
CY:ACM95,
HuthKwiakowska:LICS97,
Couvreur:2003:LPAR}).
The motivation is to 
capture the behaviors of systems with uncertainty such as
programs with unreliable channels, 
randomized algorithms, and fault-tolerant systems.
The underlying semantics for such models is often that of
a {\it Markov chain}.
In a Markov chain, each transition  is assigned a {\it probability}
by which the transition is performed from a state of the system.
In probabilistic model checking, three classes of problems are
relevant:
\begin{enumerate}[$\bullet$]
\item
The qualitative problem: check whether a certain property $\Phi$
holds with probability one (or zero).
\item
The approximate quantitative problem:
compute the probability $p$ of satisfying a given property $\Phi$ 
up-to arbitrary precision, i.e., for any pre-defined error margin
$\epsilon > 0$, compute a value $p'$ s.t. $p' \le p \le p' + \epsilon$.
\item
The exact 
quantitative problem: compute the probability $p$ of satisfying a 
given property $\Phi$ exactly and decide
exact questions, e.g., if $p \ge 0.5$.
\end{enumerate}

Recently, several attempts have been made to consider systems
which combine the above two features, i.e.,
systems which are
infinite-state and which exhibit probabilistic behavior.
For instance the works in
\cite{Rabinovich:plcs,Schnoeblen:plcs,Parosh:Alex:PLCS,BaEn:plcs,IyeNar97,ABIJ:problossy}
consider {\it Probabilistic 
Lossy Channel Systems (PLCS)}: systems consisting of finite-state
processes, which communicate through channels which are unbounded
and unreliable in the sense that they can
spontaneously lose messages.
The motivation for these works is that,
since we are dealing with unreliable communication, it is relevant
to take into consideration the probability by which messages are  lost
inside the channels.
The papers
\cite{EsparzaKuceraMayr:lics2004,EsparzaKuceraMayr:lics2005,EKM:LMCS2006,Etessami-Yannakakis:STACS05,Etessami:Yannakakis:TACAS05,Esparza:Etessami:fsttcs04K,Etessami:Yannakakis:ICALP05}
consider {\it probabilistic pushdown automata} 
(recursive state machines)
which are natural models for probabilistic sequential programs with recursive
procedures.

\paragraph{\bf Our contribution.}
Here we consider more abstract conditions on infinite Markov chains.
We show how verification problems can be solved for Markov chains with 
these conditions and that several infinite-state probabilistic process models 
satisfy them. In particular, we consider probabilistic lossy channel systems
(PLCS), probabilistic vector addition systems with states (PVASS) and 
probabilistic noisy Turing machines (PNTM).

Let $\fstates$ be a given set of target states in a Markov chain,
and $\unreachable{\fstates}$ the set of states from which $\fstates$ cannot
be reached, i.e., $\unreachable{\fstates} := \{s\ |\ s \not\transition{*}
\fstates\} = \overline{{\it Pre}^*(\fstates})$.
We call a Markov chain {\em decisive} w.r.t. a given set of
target states $\fstates$ if it almost certainly (i.e., with probability 1)
eventually reaches either 
$\fstates$ or $\unreachable{\fstates}$.
In other words, decisiveness means that if $\fstates$ is always reachable 
then it will almost certainly be reached.

While all finite Markov chains are trivially decisive 
(for every set $\fstates$), 
this also holds for several classes of infinite-state Markov chains.

It is not a meaningful question if the decisiveness property is decidable for
general Markov chains. For finite Markov chains the answer is always yes, and for
general infinite Markov chains the problem instance is not finitely given,
unless one restricts to a particular subclass. 
For some such subclasses decisiveness always holds, while for others 
(e.g., probabilistic pushdown automata (PPDA)) it is decidable (see below).

\begin{enumerate}[$\bullet$]
\item
Markov chains which contain a finite {\it attractor}.
An {\it attractor} is a set of states which is eventually reached with 
probability one from every state in the Markov chain.
Examples of Markov chains with finite attractors are all Markov chains
induced by probabilistic lossy channel systems (PLCS).

We show that infinite Markov chains which contain a finite attractor
are decisive w.r.t. {\em every} set $\fstates$.
\item
Markov chains which are {\it globally coarse}.
A Markov chain is {\it globally coarse} w.r.t. $\fstates$
if there exists some $\alpha >0$ such that, 
from every state, the probability of eventually reaching
the set $\fstates$ is either zero or $\ge \alpha$.
Global coarseness w.r.t. $\fstates$ also implies decisiveness w.r.t. $\fstates$.
We consider two probabilistic process models which induce 
globally coarse Markov chains.
\begin{enumerate}[-]
\item
Any probabilistic vector addition system with states (PVASS) 
with an upward-closed set of final states $\fstates$
induces a globally coarse Markov chain.
\item
Noisy Turing machines (NTM) have been defined by Asarin and
Collins \cite{asarin2005noisy}. These are Turing machines where the
memory tape cells are subject to `noise', i.e., random changes.
We consider probabilistic noisy Turing machines (PNTM),
a generalization of noisy Turing machines (NTM) where 
the transition steps are also chosen probabilistically.
Probabilistic noisy Turing machines induce 
globally coarse Markov chains w.r.t. {\em every} set $\fstates$ defined by 
a set of control-states.
\end{enumerate}
\item
Another subclass of infinite Markov chains are those induced by
probabilistic pushdown automata (PPDA; also called recursive state machines)
\cite{EsparzaKuceraMayr:lics2004,EsparzaKuceraMayr:lics2005,EKM:LMCS2006,Etessami-Yannakakis:STACS05,Etessami:Yannakakis:TACAS05,Esparza:Etessami:fsttcs04K,Etessami:Yannakakis:ICALP05}.
These infinite Markov chains are not decisive in general. However, it follows
directly from the results in \cite{EKM:LMCS2006} that
decisiveness is decidable for PPDA, provided that the set of final states 
$\fstates$ is effectively regular.
\end{enumerate}

The focus of this paper are the classes PLCS, PVASS and PNTM, not PPDA.
We strive to be as general as possible and use only the weak condition
of decisiveness. We do not advocate the use of our algorithms for PPDA,
even for those instances which are decisive. Since PPDA is a special class
with a particular structure, specialized algorithms like those
described in
\cite{EsparzaKuceraMayr:lics2004,EsparzaKuceraMayr:lics2005,EKM:LMCS2006,Etessami-Yannakakis:STACS05,Etessami:Yannakakis:TACAS05,Esparza:Etessami:fsttcs04K,Etessami:Yannakakis:ICALP05}
are more suitable for it. 
However, we show in Section~\ref{sec:exact_quant_reach} that the techniques
used for analyzing PPDA cannot be applied to PLCS, PVASS or PNTM.

We consider both qualitative and quantitative analysis for 
decisive Markov chains.
The main contributions of the paper are the following.

\begin{enumerate}[$\bullet$]
\item
The qualitative reachability problem, i.e., the question if $\fstates$ is
reached with probability 1 (or 0).
For decisive Markov chains, this problem
is equivalent to a question about the underlying 
(non-probabilistic) transition system. 

For PVASS, the decidability of this question depends on the set of
target states $\fstates$.
It is decidable if $\fstates$ is defined by a set of control-states,
but undecidable if $\fstates$ is a more general upward-closed set of
configurations. 
This is in contrast to all known decidability results for other models
such as non-probabilistic VASS, and PLCS, where the two problems
can effectively be reduced to each other.

For both PLCS and PNTM, the qualitative reachability problem is generally decidable.
In particular for PLCS, although this was already shown in \cite{Parosh:Alex:PLCS,Schnoeblen:plcs},
our construction is more abstract and simpler.
In particular, our algorithm does not require explicit construction
of the attractor as in \cite{Parosh:Alex:PLCS,Schnoeblen:plcs}.
\item
The qualitative repeated reachability problem. 

If a Markov chain is decisive w.r.t. $\fstates$ then the question whether
$\fstates$ will be visited infinitely often with probability 1 is equivalent
to a simple question about the underlying transition graph, which is decidable 
for PVASS, PLCS and PNTM. For PVASS, the decidability of probabilistic 
repeated reachability is surprising, given the undecidability of probabilistic 
simple reachability above.

If a Markov chain is decisive w.r.t.\ both $\fstates$ and
$\unreachable{\fstates}$ then the question whether
$\fstates$ will be visited infinitely often with probability 0 is equivalent
to another question about the underlying transition graph.
The precondition holds for all Markov chains with a finite attractor (such a
PLCS) since they are decisive w.r.t. every set, and the question is decidable
for PLCS.\/ For PNTM, we show that if $\fstates$ is defined by a set of 
control-states then so is $\unreachable{\fstates}$. Since PNTM induce 
globally coarse Markov chains w.r.t. any set defined by control-states, 
the question is also decidable.

However, for PVASS, decisiveness w.r.t. $\fstates$ does not generally imply 
decisiveness w.r.t. $\unreachable{\fstates}$ and thus our algorithm
is not always applicable. For PVASS, decidability of the question
whether $\fstates$ is visited infinitely often with probability 0 is 
an open problem.

{\small
\begin{table}
\begin{minipage}{0.97\columnwidth}\centering
\begin{tabular}{|p{4.3cm}|p{3cm}|p{3cm}|p{3cm}|} \cline{2-4}
 \multicolumn {1}{l|}{}    & {\bf PLCS} & {\bf PVASS} & {\bf PNTM} \\
\hline
Approximate $\probms{\mchain}(\initstate\models\eventually\fstates)$
&
Solvable when $\fstates$ 

is effectively 

representable;

see Theorem~\ref{quantitative:plcs:theorem} 
&
Solvable when $\fstates$

is upward-closed;

\phantom{xxx}

see Theorem~\ref{quantitative:vass:theorem} 
&
Solvable when $\fstates$

is defined by 

control-states;

see Theorem~\ref{quantitative:ntm:theorem} \\
\cline{1-4}
Approximate $\probms{\mchain}(\initstate\models\always\eventually\fstates)$
&
Solvable when $\fstates$

is effectively

representable;

see Theorem~\ref{quantitative:repeated:plcs:theorem} 
&
Open problem
&
Solvable when $\fstates$

is defined by

control-states;

see Theorem~\ref{quantitative:repeated:ntm:theorem} \\
\cline{1-4}
Compute the exact

$\probms{\mchain}(\initstate\models\eventually\fstates)$, or
$\probms{\mchain}(\initstate\models\always\eventually\fstates)$%
~\footnote{All results here concern the effective expressibility of 
 the probability in Tarski-algebra.}
&\raggedright
Not constructible when $\fstates$ is defined by control-states;

see Theorem~\ref{thm:plcs_not_constructible}/ Remark~\ref{rem:rep_reach_nonexp} 
&\raggedright
Not constructible when $\fstates$ is defined by control-states;

see Theorem~\ref{thm:not_constructible}/ Remark~\ref{rem:rep_reach_nonexp} 
&
Not constructible

when $\fstates$ is defined

by control-states;

see Theorem~\ref{thm:ntm_not_constructible}/ 

Remark~\ref{rem:rep_reach_nonexp} \\
\cline{1-4}
\hline
\end{tabular}
\caption{Computability results for quantitative problems}
\label{tab:quant_res}
\end{minipage}
\end{table}
}

\item
To approximate the probability of eventually reaching $\fstates$, we
recall an algorithm from \cite{IyeNar97} which was also used in
\cite{Rabinovich:plcs} for PLCS.\/ We show that 
the algorithm can be used to solve the problem
for all decisive Markov chains (in particular also for both PVASS and PNTM).

Furthermore, we show that a minor modification of the
algorithm yields an algorithm for approximating the probability
of visiting $\fstates$ infinitely often for all Markov chains which are decisive w.r.t.
$\fstates$ and $\unreachable{\fstates}$. 
In particular this works for all Markov chains with
a finite attractor, such as PLCS.\/ This is a more abstract, 
general and simpler
solution than the result for PLCS in \cite{Rabinovich:plcs}. However, 
it does not yield precise complexity bounds as \cite{Rabinovich:plcs}. 
\item
The question if the exact probability of (either eventually, or infinitely
often) reaching $\fstates$ in PLCS is expressible by standard mathematical functions 
was stated as an open problem in \cite{Rabinovich:plcs}.
We provide a partial answer by showing that
for PVASS, PLCS and (P)NTM, this probability
cannot be effectively expressed (in a uniform way) in Tarski-algebra, 
the first-order theory of the reals
$(\R,+,*,\le)$. (By `in a uniform way' we mean that quantitative parameters in
the system should be reflected directly by constants in the Tarski-algebra-formula.)
This is in contrast to the situation for 
probabilistic pushdown automata for which these probabilities can be effectively
expressed, in a uniform way, 
in $(\R,+,*,\le)$ \cite{EsparzaKuceraMayr:lics2004,EKM:LMCS2006,Etessami-Yannakakis:STACS05,Esparza:Etessami:fsttcs04K}.
\end{enumerate}

{\small
\begin{table}
\begin{minipage}{0.97\columnwidth}\centering
\begin{tabular}{|p{4.3cm}|p{3cm}|p{3cm}|p{3cm}|} \cline{2-4}
 \multicolumn {1}{l|}{}    & {\bf PLCS} & {\bf PVASS} & {\bf PNTM} \\
\hline
Approximate $\probms{\mchain}(\initstate\models\eventually\fstates)$
&
Solvable when $\fstates$ 

is effectively 

representable;

see Theorem~\ref{quantitative:plcs:theorem} 
&
Solvable when $\fstates$

is upward-closed;

\phantom{xxx}

see Theorem~\ref{quantitative:vass:theorem} 
&
Solvable when $\fstates$

is defined by 

control-states;

see Theorem~\ref{quantitative:ntm:theorem} \\
\cline{1-4}
Approximate $\probms{\mchain}(\initstate\models\always\eventually\fstates)$
&
Solvable when $\fstates$

is effectively

representable;

see Theorem~\ref{quantitative:repeated:plcs:theorem} 
&
Open problem
&
Solvable when $\fstates$

is defined by

control-states;

see Theorem~\ref{quantitative:repeated:ntm:theorem} \\
\cline{1-4}
Compute the exact

$\probms{\mchain}(\initstate\models\eventually\fstates)$, or
$\probms{\mchain}(\initstate\models\always\eventually\fstates)$%
~\footnote{All results here concern the effective expressibility of 
 the probability in Tarski-algebra.}
&\raggedright
Not constructible when $\fstates$ is defined by control-states;

see Theorem~\ref{thm:plcs_not_constructible}/ Remark~\ref{rem:rep_reach_nonexp} 
&\raggedright
Not constructible when $\fstates$ is defined by control-states;

see Theorem~\ref{thm:not_constructible}/ Remark~\ref{rem:rep_reach_nonexp} 
&
Not constructible

when $\fstates$ is defined

by control-states;

see Theorem~\ref{thm:ntm_not_constructible}/ 

Remark~\ref{rem:rep_reach_nonexp} \\
\cline{1-4}
\hline
\end{tabular}
\caption{Computability results for quantitative problems}
\label{tab:quant_res}
\end{minipage}
\end{table}
}

%
%
%
%
%

\section{Transition Systems and Markov Chains}
\label{mchain:section}

\noindent
We introduce some basic concepts for transition systems and
Markov chains. Let $\nat$ and $\rat_{\ge 0}$ denote the set of 
natural numbers (including 0)
and non-negative rational numbers, respectively.

\subsection{Transition Systems}

\noindent
A {\it transition system $\tssystem$} is a tuple $\tstuple$ where
$\states$ is a (potentially) infinite set of states, and
$\transition{}$ is a binary relation on $\states$.
We write $\state\transition{}\state'$ for
$\tuple{\state,\state'}\in\transition{}$ and 
let $\post(\state) :=\setcomp{\state'}{\state\transition{}\state'}$.
A {\it run $\run$ (from $\state_0$)} of $\tssystem$ is an infinite sequence
$\state_0\state_1\ldots$ of states such that $\state_i\transition{}\state_{i+1}$
for $i\geq 0$. We use $\run(i)$ to denote $\state_i$ and say that $\run$ is an 
$\state$-run if $\run(0)=\state$.
A {\it path} is a finite prefix of a run.
We assume familiarity with the syntax and semantics of the temporal logic
\ctlstar{} \cite{CGP:book}.
We use $\left(\state\models\phi\right)$ to denote the 
set of $\state$-runs that satisfy the \ctlstar{} path-formula $\phi$.
For $\state\in\states$ and $\qstates\subseteq\states$, we say that $\qstates$ is 
{\it reachable} from $\state$ if $\state\models\ef\qstates$.
For 
$\qstates_1,\qstates_2\subseteq\states$, we use
$\qstates_1\before\qstates_2$ 
to denote the CTL formula 
$\exists\left(\neg\qstates_2\,\until\,\qstates_1\right)$, i.e., there exists
a run which reaches a state in $\qstates_1$ without having 
previously passed through any state in $\qstates_2$.
Given a set of states $\fstates\subseteq\states$, 
we define ${\it Pre}^*(\fstates) := \{s'\ |\ \exists s\in\fstates:\,s' \transition{*} s\}$
as the set of its predecessors. Furthermore, let
$\unreachable{\fstates} := \overline{{\it Pre}^*(\fstates)} 
= \setcomp{\state}{\state\not\models\ef\fstates}$, 
the set of states from which $\fstates$ is not reachable.
For $\state\in\states$ and $\fstates\subseteq\states$, we define the {\it distance} $\dist{\fstates}(\state)$
of $\state$ to $\fstates$ to be the minimal natural number $n$ 
with $\state\transition{n}\fstates$.
In other words, $\dist{\fstates}(\state)$
is the length of the shortest path leading 
from $\state$ to $\fstates$.
In case $\state\in\unreachable{\fstates}$, we define $\dist{\fstates}(\state)=\infty$.
A transition system $\tssystem$ is said to be of {\it span $N$} with respect 
to a given set $\fstates$ if for each $\state\in\states$
we either have $\dist{\fstates}(\state)\leq N$ or $\dist{\fstates}(\state)=\infty$.
We say that $\tssystem$ is {\it finitely spanning} with respect to a given set 
$\fstates$ if $\tssystem$ is of span $N$ w.r.t. $\fstates$ for some $N\geq 0$.
A transition system $\tssystem=\tstuple$  
is said to be {\it effective} w.r.t. a given set $\fstates$ if for each 
$\state\in\states$, we can 
(1) compute elements of the set $\post(\state)$
(notice that this implies that $\tssystem$ is finitely branching); and
(2) check whether $\state\models\ef \fstates$.

\subsection{Markov Chains}

\noindent
A {\it Markov chain} $\mchain$ is a tuple $\mctuple$
where $\states$ is a (potentially infinite) set of {\it states}, and
$\probability: \states\times\states \rightarrow [0,1]$, 
such that $\sum_{\state'\in\states}\probability(\state,\state')=1$,
for each $\state\in\states$.
A Markov chain induces a transition system, where the transition
relation consists of pairs of states related by positive probabilities.
In this manner, concepts defined for transition systems can be lifted
to Markov chains.
For instance, for a Markov chain $\mchain$, a run of $\mchain$ is a run
in the underlying transition system, and
$\mchain$ is {\it finitely spanning} w.r.t. given set $\fstates$
if the underlying transition system is finitely spanning w.r.t. $\fstates$, etc.

Consider a state $\state_0$ of a Markov chain $\mchain=\mctuple$.
On the sets of $\state_0$-runs, the probability space $(\Om,\De,\probms{\mchain})$ 
is defined as follows (see also \cite{KSK:book}):
$\Om=\state_0\states^{\om}$ is the set
of all infinite sequences of states starting from
$\state_0$, ${\De}$ is the $\sigma$-algebra generated by the basic cylindric sets
$D_u=u\states^{\om}$, for every $u\in \state_0 S^*$, and the probability measure $\probms{\mchain}$
is defined by $\probms{\mchain}(D_u)=\prod_{i=0,...,n-1}P(s_i,s_{i+1})$ where
$u=s_0s_1...s_{n}$; this measure is extended in a unique way to the elements of the 
$\sigma$-algebra generated by the basic cylindric sets.

Given a \ctlstar{} path-formula $\phi$,
we use $(\state\models\phi)$ 
to denote the set of $\state$-runs 
that satisfy $\phi$.
We use 
$\probms{\mchain}\left(\state\models\phi\right)$ to denote the measure 
of the set of $\state$-runs $\left(\state\models\phi\right)$ (which is measurable
by \cite{Vardi:probabilistic}).
For instance, given a set $\fstates\subseteq\states$, 
$\probms{\mchain}\left(\state\models\eventually\fstates\right)$
is the measure of $\state$-runs which eventually reach $\fstates$.
In other words, it is the probability by which $\state$ satisfies
$\eventually\fstates$.
We say that {\em almost all} runs of a Markov chain satisfy a given
property $\phi$ if $\probms{\mchain}\left(\state\models\phi\right)=1$.
In this case one says that $(\state\models\phi)$ holds
{\em almost certainly}.

\section{Classes of Markov Chains}
\label{classes:section}

\noindent
In this section we define several abstract properties of infinite-state
Markov chains: decisiveness, the existence of a finite
attractor, and global coarseness. We show that both the existence of a finite
attractor and global coarseness imply decisiveness.
In particular, all three properties hold trivially for finite Markov chains.

In the rest of this section, we assume a Markov chain $\mchain=\mctuple$.

\subsection{Decisive Markov Chains}

\begin{defi}\label{def:decisive}
Given a Markov chain $\mchain=\mctuple$ and a set of states 
$\fstates\subseteq\states$, we say that $\mchain$ is {\it decisive} w.r.t. $\fstates$ iff
$\probms{\mchain}(\state\models\eventually\fstates\vee\eventually\unreachable{\fstates})=1$, 
for each $\state\in\states$. 

In other words, the set of runs, along which $\fstates$ is always
reachable but which never reach $\fstates$, is almost empty (i.e.,
has probability measure zero).

Similarly, we say that $\mchain$ is {\it strongly decisive} w.r.t.\ $\fstates$ if 
$\probms{\mchain}(\state\models\eventually\unreachable{\fstates}\vee\always\eventually\fstates)=1$. 
Intuitively, this means that
the set of runs along which $\fstates$ is always reachable and which visit
$\fstates$ only finitely many times is almost empty.
\end{defi}

\begin{lem}\label{lem:decisive}
Given a Markov chain $\mchain=\mctuple$ and a set $\fstates\subseteq\states$, $\mchain$ 
is decisive w.r.t. $\fstates$ iff it is strongly decisive w.r.t. $\fstates$.
\end{lem}
\begin{proof}
Given a Markov chain $\mchain=\mctuple$ and a set $\fstates\subseteq\states$, we want 
to prove that $\forall \state\in\states
,\probms{\mchain}(\state\models\eventually\fstates\vee\eventually\unreachable{\fstates})=1\Longleftrightarrow\forall\state\in\states
,\probms{\mchain}(\state\models\eventually\unreachable{\fstates}\vee\always\eventually\fstates)=1$.
This is equivalent to proving that 
\[
\forall \state\in\states
,\probms{\mchain}(\state\models\always\neg\fstates\wedge\always\neg\unreachable{\fstates})=0\Longleftrightarrow\forall\state\in\states
,\probms{\mchain}(\state\models\eventually\always\neg\fstates\wedge\always\neg\unreachable{\fstates})=0.
\]
Let $U$ be a set of sequences of states. $U$ is called {\it proper} if no sequence in $U$ 
is a prefix of another sequence in $U$. If all sequences in $U$ are finite and 
start at the same state, we define $P(U):=\probms{\mchain}(D_U)$ where 
$D_U=\setcomp{u\states^{\om}}{u\in U}$. 
Given a proper set $U$ of finite sequences (namely paths) ending all in the same 
state $\state_c$ and a proper set $V$ of possibly infinite sequences (runs) starting all from 
$\state_c$, we define $U\bullet V$ to be the set of all sequences 
$u\state_c v$ where $u\state_c\in U$ and $\state_c v\in V$.

We now prove both implications of the required equivalence above.

($\Longleftarrow$) Observe that $(\state\models\eventually\always\neg\fstates)$ is the set of 
$\state$-runs visiting $\fstates$ only finitely many times.
In particular, the set of $\state$-runs which never visit $\fstates$ is included in that set.
This gives $(\state\models\always\neg\fstates)\subseteq(\state\models\eventually\always\neg\fstates)$.
By intersection with $(\state\models\always\neg\unreachable{\fstates})$, 
the set of $\state$-runs which never visit $\unreachable{\fstates}$, 
we obtain 
$(\state\models\always\neg\fstates\wedge\always\neg\unreachable{\fstates})\subseteq(\state\models\eventually\always\neg\fstates\wedge\always\neg\unreachable{\fstates})$. 

By definition of the probability measure, we obtain 
$\probms{\mchain}(\state\models\always\neg\fstates\wedge\always\neg\unreachable{\fstates})\le
\probms{\mchain}(\state\models\eventually\always\neg\fstates\wedge\always\neg\unreachable{\fstates})=0$
for any $\state\in\states$, where the last equality follows from the assumption.

($\Longrightarrow$) Given a state $\state\in\states$, we define the following sets of paths:
\[\pAthes^{\state}_{ir} 
:= \{\pAth\,|\,\pAth=\state(\overline{\fstates}^* \fstates)^ir\}
\quad\hbox{where}\quad
i\ge0\quad\hbox{and}\quad r\in\overline{\fstates}\ .
\]
%
Now, consider the following sets of runs:
{\small
\[\eqalign{
 \forall i\ge0\,\Gamma^{\state}_i 
&:= \textstyle\bigcup_{r\in\overline{\fstates}}\Gamma^{\state}_{ir}
  \quad\hbox{where}\quad 
  \forall r\in\overline{\fstates}\,
  \Gamma^{\state}_{ir} := 
  \pAthes^{\state}_{ir}\bullet 
  (r\models\always\neg\fstates\wedge\always\neg\unreachable{\fstates})\cr
 \forall i\ge0\,\Delta^{\state}_i 
&:= \textstyle\bigcup_{r\in\overline{\fstates}}\Delta^{\state}_{ir}
  \quad\hbox{where}\quad
  \forall r\in\overline{\fstates}\,
  \Delta^{\state}_{ir} := 
  \pAthes^{\state}_{ir}\bullet (r\states^{\om} )\ .\cr
}
\]}
Intuitively, $\Delta^{\state}_i$ is the set of $\state$-runs which revisit $\fstates$ 
at least $i$ times while $\Gamma^{\state}_i$ is the set of all $\state$-runs 
which revisit $\fstates$ exactly $i$ times and then never visit neither $\fstates$ nor 
$\unreachable{\fstates}$. Observe that for $i=0$, $\Gamma^{\state}_0=(\state\models\always\neg\fstates\wedge\always\neg\unreachable{\fstates})$.
It is straightforward to check that: 
{\small
\begin{enumerate}[(1)]
\item \label{state_disjoint}
$(\forall i\in \nat)(\forall r_1,r_2\in\overline{\fstates}\wedge r_1\neq r_2)(\Gamma^{\state}_{i r_1}\cap\Gamma^{\state}_{i r_2}=\emptyset)$
\item \label{i_disjoint}
$(\forall i,j\in\nat\wedge i\neq j)(\Gamma^{\state}_i\cap\Gamma^{\state}_j=\emptyset)$
\item \label{gamma_prob}
$(\forall i\in\nat)(\forall r\in\overline{\fstates})(\probms{\mchain}(\Gamma^{\state}_{i r})=P(\pAthes^{\state}_{i r})\probms{\mchain}(r\models\always\neg\fstates\wedge\always\neg\unreachable{\fstates}))$
\end{enumerate}
}
Therefore, it follows that for all $i\in\nat$ 
{\small
\begin{align*}
\probms{\mchain}(\Gamma^{\state}_{i})&=\sum_{r\in\overline{\fstates}}\probms{\mchain}(\Gamma^{\state}_{i r})=\sum_{r\in\overline{\fstates}}P(\pAthes^{\state}_{i r})\probms{\mchain}(r\models\always\neg\fstates\wedge\always\neg\unreachable{\fstates})=0 
\end{align*}
}
where the first equality holds by (\ref{state_disjoint}).
The second equality follows from (\ref{gamma_prob}), and the last from the fact that $\mchain$ is decisive w.r.t. $\fstates$; i.e., for all $r\in\states$, $\probms{\mchain}(r\models\always\neg\fstates\wedge\always\neg\unreachable{\fstates})=0$.

Observe that $\bigcap_{i=0}^{\infty} \Delta^{\state}_{i}\subseteq (\state\models\always\eventually\fstates)$.
Therefore, $(\state\models\eventually\always\neg\fstates)\subseteq\bigcup_{i=0}^{\infty}\neg\Delta^{\state}_{i}$ where for all $i\ge0$, $\neg\Delta^{\state}_{i}$ is the set of $\state$-runs revisiting $\fstates$ at most $i-1$ times. 
For all $i\ge 0$, we have $(\neg\Delta^{\state}_{i}\cap(\state\models\always\neg\unreachable{\fstates}))\subseteq\bigcup_{j=0}^{i-1}\Gamma^{\state}_{j}$.
By using this inclusion, property~(\ref{i_disjoint}), and the fact that
$\Gamma^{\state}_{i}$ has measure zero, we obtain {\small $\probms{\mchain}(\state\models\eventually\always\neg\fstates\wedge\always\neg\unreachable{\fstates})\leq\probms{\mchain}\left(\bigcup_{i=0}^{\infty}\Gamma^{\state}_{i}\right)=\sum_{i=0}^{\infty}\probms{\mchain}(\Gamma^{\state}_{i})=0$}.
\end{proof}

\subsection{Markov Chains with a Finite Attractor}

\begin{defi}\label{def:attractor}
Given a Markov chain $\mchain=\mctuple$, a set $\attractor\subseteq\states$ 
is said to be an {\it attractor},
if for each $\state\in\states$, we have 
$\probms{\mchain}\left(\state\models\eventually\attractor\right)=1$, i.e.,
the set $\attractor$ is reached from $\state$ with probability one.
\end{defi}

\begin{lem}
\label{lem:attractor_decisive}
A Markov chain $\mchain$ which has a finite attractor is decisive w.r.t. every
set $\fstates \subseteq \states$.
\end{lem}
\begin{proof}
Fix a Markov chain $\mchain=\mctuple$ that has a finite attractor 
$\attractor$, a state $\state$ and a set $\fstates\subseteq\states$.
Recall that $(\state\models\always\neg\fstates\wedge\always\neg\unreachable{\fstates})$ 
denotes the set of $\state$-runs which neither visit $\fstates$ nor $\unreachable{\fstates}$.
All states $\state'$ visited by runs in $(\state\models\always\neg\fstates\wedge\always\neg\unreachable{\fstates})$ 
satisfy $\state'\models\exists\eventually\fstates$, since $\state' \notin \unreachable{\fstates}$.
In particular this holds for the finitely many different $\state''\in\attractor$ visited by those runs.
Let $\attractor' \subseteq \attractor$ denote the set of states from the attractor,
visited by runs in  $(\state\models\always\neg\fstates\wedge\always\neg\unreachable{\fstates})$. 
For every $\state'' \in \attractor'$ we define
$\alpha_{\state''} := \probms{\mchain}(\state'' \models
\eventually\fstates)$, and obtain $\alpha_{\state''} > 0$.
By definition of an attractor, we obtain that $\attractor'$ is not empty.
By finiteness of $\attractor$ (and thus $\attractor'$), 
it follows that $\alpha := \min_{\state'' \in A'}
\alpha_{\state''} > 0$. 
Almost every run must visit $\attractor$
infinitely often, and only states in $\attractor'$ are visited by runs in
$(\state\models\always\neg\fstates\wedge\always\neg\unreachable{\fstates})$. 
Thus, 
$\probms{\mchain}(\state\models\always\neg\fstates\wedge\always\neg\unreachable{\fstates}) 
\le (1-\alpha)^\infty = 0$.
Finally, we obtain for each 
$\state \in \states$, $\probms{\mchain}(\state\models\eventually\fstates\vee\eventually\unreachable{\fstates})=1-\probms{\mchain}(\state\models\always\neg\fstates\wedge\always\neg\unreachable{\fstates})=1$.
\end{proof}

\subsection{Globally Coarse Markov Chains}

\begin{defi}\label{def:coarse}
A state $\state$ is said to be of {\it coarseness $\beta$} if for each $\state'\in\states$,
$\probability(\state,\state')>0$ implies $\probability(\state,\state')\geq \beta$.
A Markov chain $\mchain=\mctuple$ is said to be of 
{\it coarseness $\beta$} if each $\state\in\states$ is of coarseness $\beta$.
We say that $\mchain$ is {\it coarse} if $\mchain$ is of coarseness $\beta$, 
for some $\beta>0$.
Notice that if $\mchain$ is coarse then the underlying transition system is
finitely branching; however, the converse is not necessarily true.
Given a Markov chain $\mchain=\mctuple$ and a set $\fstates\subseteq\states$. 
We say that a Markov chain $\mchain=\mctuple$ is {\em globally coarse} w.r.t. $\fstates$
if there exists some $\alpha > 0$ 
s.t. $\forall \state\in \states.\, (\state \transition{*}\fstates)
\Rightarrow \probms{\mchain}(\state \models \eventually\fstates) \ge \alpha$.
\end{defi}

\begin{lem}\label{lem:globally_coarse}
If a Markov chain is coarse and finitely spanning w.r.t.\ a set $\fstates$ then
it is globally coarse w.r.t.\ $\fstates$.
\end{lem}
\begin{proof}
If a Markov chain is coarse (of coarseness $\beta >0$) and finitely
spanning w.r.t.\ a given set $\fstates$ (of span $N$) then it is globally coarse 
w.r.t.\ the same set $\fstates$ (define $\alpha:=\beta^N$).
\end{proof}

\begin{lem}
\label{lem:gcoarse_decisive}
Given a Markov chain $\mchain$ and a set $\fstates$ such that $\mchain$ is 
globally coarse w.r.t.\ $\fstates$, then $\mchain$ is decisive w.r.t.\ $\fstates$.
\end{lem}
\begin{proof}
Assume a Markov chain $\mchain=\mctuple$, a state $\state$ and a set $\fstates\subseteq\states$ such that $\mchain$ is 
globally coarse w.r.t.\ $\fstates$.
All states $\state'$ visited by runs in $(\state\models\always\neg\fstates\wedge\always\neg\unreachable{\fstates})$ 
satisfy $\state' \models \exists\eventually\fstates$, because $\state' \notin \unreachable{\fstates}$. 
Since $\mchain$ is globally coarse w.r.t.\ $\fstates$ there exists some universal constant
 $\alpha >0$ s.t. $\probms{\mchain}\left(\state' \models \eventually\fstates\right) \ge
 \alpha$ for any $\state'$ which is visited by those runs. 
Therefore, $\probms{\mchain}(\state\models\always\neg\fstates\wedge\always\neg\unreachable{\fstates}) \le (1-\alpha)^\infty=0$.
Finally, we obtain for each 
$\state\in\states$, $\probms{\mchain}(\state\models\eventually\fstates\vee\eventually\unreachable{\fstates})=1-\probms{\mchain}(\state\models\always\neg\fstates\wedge\always\neg\unreachable{\fstates})=1$.
\end{proof}

\section{System Models and their Properties}
\label{sec:models}
\noindent
We define three classes of infinite-state probabilistic system models 
and describe the induced Markov chains.

\subsection{Vector Addition Systems}

\noindent
A {\it Vector Addition System with States (VASS)} consists of a finite-state process
operating on a finite set of unbounded variables each of which ranges over $\nat$.
Formally, a {\it VASS $\vass$} is a tuple
$\vasstuple$, where $\vassstates$ is a finite set of {\it control-states},
$\vars$ is a finite set of {\it variables},
and $\vasstransitions$ is a set of {\it transitions}
each of the form $\tuple{\vassstate_1,\op,\vassstate_2}$, where
$\vassstate_1,\vassstate_2\in\vassstates$, and
$\op$ is a mapping from $\vars$ to the set $\set{-1,0,1}$.
A {\it (global) state} $\state$ is of the form $\tuple{\vassstate,\varstate}$
where $\vassstate\in\vassstates$ and $\varstate$ is a mapping from
$\vars$ to $\nat$.
%

%
We use $\vassstate$ and $\vassstates$ to range over control-states and sets
of control-states, respectively.
On the other hand, we use
$\state$ and $\states$ to range over states and sets of states of
the induced transition system (states of the transition system are global states
of the VASS).

For $\varstate_1,\varstate_2\in(\vars\mapsto\nat)$, we
use $\varstate_1\preceq\varstate_2$ to denote that
$\varstate_1(\var)\leq\varstate_2(\var)$ for each
$\var\in\vars$.
We extend $\preceq$ to a relation on
$\vassstates\times\left(\vars\mapsto\nat\right)$,
where $\tuple{\vassstate_1,\varstate_1}\preceq\tuple{\vassstate_2,\varstate_2}$
iff $\vassstate_1=\vassstate_2$ and $\varstate_1\preceq\varstate_2$.

A set of global states $\fstates$ is upward-closed w.r.t.\ $\preceq$
iff 
\[
\tuple{\vassstate_1,\varstate_1} \in \fstates\,\wedge\,
\tuple{\vassstate_1,\varstate_1}\preceq\tuple{\vassstate_2,\varstate_2}
\ \Rightarrow
\ \tuple{\vassstate_2,\varstate_2} \in \fstates
\]
Analogously, a set of global states $\fstates$ is downward-closed w.r.t.\ $\preceq$
iff
\[
\tuple{\vassstate_2,\varstate_2} \in \fstates\,\wedge\,
\tuple{\vassstate_1,\varstate_1}\preceq\tuple{\vassstate_2,\varstate_2}
\ \Rightarrow
\ \tuple{\vassstate_1,\varstate_1} \in \fstates
\]
The complement of an upward-closed set is downward-closed and vice-versa.

For $\qvassstates\subseteq\vassstates$, we define
a {\it $\qvassstates$-state} to be a state of the form
$\tuple{\vassstate,\varstate}$ where $\vassstate\in\qvassstates$.
Notice that, for any $\qvassstates\subseteq\vassstates$,
the set of $\qvassstates$-states is  upward-closed and 
downward-closed with respect to $\preceq$.

It follows from Dickson's Lemma~\cite{Dickson:lemma}
that every infinite set of VASS configurations has only finitely many minimal
elements w.r.t.\ $\preceq$. When we speak of an upward-closed set of VASS configurations,
we assume that it is represented by its finitely many minimal elements.

A transition $\vasstransition=\tuple{\vassstate_1,\op,\vassstate_2}$ is said to be
{\it enabled} at $\tuple{\vassstate_1,\varstate_1}$ if
$\varstate_1(\var)+\op(\var)\geq 0$ for each $\var\in\vars$.
We define $\enabled(\vassstate,\varstate)=
\setcomp{\vasstransition}{\mbox{$\vasstransition$ is enabled at $\tuple{\vassstate,\varstate}$}}$.
In case $\vasstransition=\tuple{\vassstate_1,\op,\vassstate_2}$ is enabled
at $\tuple{\vassstate_1,\varstate_1}$, we define
$\vasstransition(\vassstate_1,\varstate_1)$ to be 
$\tuple{\vassstate_2,\varstate_2}$
where
$\varstate_2(\var)=\varstate_1(\var)+\op(\var)$ for each $\var\in\vars$.
The VASS $\vass$\ignore{, together with a set $\fstates$ of global states,
induces a transition system
$\tuple{\states,\transition{},\fstates}$} induces a transition system $\tstuple$, where
$\states$ is the set of states, i.e.,
$\states=\left(\vassstates\times(\vars\mapsto\nat)\right)$, 
and
$\tuple{\vassstate_1,\varstate_1}\transition{}\tuple{\vassstate_2,\varstate_2}$
iff there is a $\vasstransition\in\vasstransitions$ with 
$\tuple{\vassstate_2,\varstate_2}=\vasstransition(\vassstate_1,\varstate_1)$.
In the sequel, we assume, without loss of generality,
that for all $\tuple{\vassstate,\varstate}$, the
set $\enabled(\vassstate,\varstate)$ is not empty, i.e.,
there is no deadlock.
This can be guaranteed 
by requiring that from each control-state there is a self-loop
not changing the values of the variables.

VASS are expressively equivalent to Petri nets \cite{Peterson:PN}.
The only difference is that VASS explicitly mention the finite control as 
something separate, while Petri nets encode it as another variable in the
vector. The reachability problem for Petri nets/VASS is decidable 
\cite{Mayr:SIAM84} and a useful extension of this result has been shown 
by Jan\v{c}ar \cite{Jancar:TCS}. In our VASS terminology this result can be
stated as follows.

\begin{thm}\label{thm:jancar}(\cite{Jancar:TCS})
Let $\vasstuple$ be a VASS with control-states $\vassstates = \{s_1,\dots,s_j\}$
and variables $\vars=\{x_1,\dots,x_n\}$. A simple constraint logic is used to
describe properties of global states
$(s,x_1,\dots,x_n)$. Any formula $\Phi$ in this logic is a boolean
combination of predicates of the following form: $s=s_i$ (the control-state is
$s_i$), $x_i \ge k$, or $x_i \le k$, where $k \in \nat$.

In particular, all upward-closed sets of VASS states can be described in this
logic. It suffices to specify that the global state must be larger or equal
(in every variable) than some of the (finitely many) minimal elements of the 
set. Since this constraint logic is closed under negation, all downward-closed
sets can also be described in it.

Given an initial global state $\tuple{\vassstate,\varstate}$, and a constraint logic formula $\Phi$,
it is decidable if there exists a reachable state that satisfies $\Phi$.
\end{thm}

{\bf\noindent  Probabilistic VASS.}
A {\it probabilistic VASS (PVASS) $\vass$} is of the form
$\pvasstuple$,
where $\vasstuple$ is a VASS and $\vassweight$ is a mapping from 
$\vasstransitions$ to the set of positive natural numbers.
Intuitively, we derive a Markov chain from $\pvass$ by assigning
probabilities to the transitions of the underlying transition system.
The probability of performing a transition $\vasstransition$ from
a state $\tuple{\vassstate,\varstate}$ is determined by the
weight $\vassweight(\vasstransition)$ of $\vasstransition$ compared
to the weights of the other transitions which are enabled at
$\tuple{\vassstate,\varstate}$.
We define
$\vassweight(\vassstate,\varstate)=
\sum_{\vasstransition\in\enabled(\vassstate,\varstate)}\vassweight(\vasstransition)$.
The PVASS $\pvass$\ignore{,
together with a set $\fstates$ of global states,}
induces a Markov chain
$\mctuple$, where $\states$ is defined as for a VASS,
and
\[
\probability\left(\tuple{\vassstate_1,\varstate_1},\tuple{\vassstate_2,\varstate_2}\right)
\;\;=\;\;
\frac{\sum\setcomp{\vassweight(\vasstransition)}
{\vasstransition(\vassstate_1,\varstate_1)=
\tuple{\vassstate_2,\varstate_2}}}{\vassweight(\vassstate_1,\varstate_1)}
\]
Notice that this is well-defined 
since $\vassweight(\vassstate_1,\varstate_1)>0$
by the assumption that there are no deadlock states.

\begin{rem}\label{rem:pvass}
Coarseness of Markov chains induced by PVASS follows immediately from
the definitions. 
It follows from results in \cite{Parosh:Bengt:Karlis:Tsay:general:IC} (Section
4 and 7.2) that each Markov chain
induced by a PVASS is effective and finitely spanning w.r.t.\ any 
upward-closed set of final markings $\fstates$. VASS induce well-structured
systems in the sense of \cite{Parosh:Bengt:Karlis:Tsay:general:IC} and
the computation of the set of predecessors of an ideal (here this means an 
upward-closed set) converges after some finite number $k$ of steps. This 
yields the finite span $k$ w.r.t.\ $\fstates$ of our Markov chain derived 
from a PVASS. 
\end{rem}

By applying Remark~\ref{rem:pvass} and 
Lemma~\ref{lem:globally_coarse} we obtain the following theorem.

\begin{thm}
\label{vass:theorem}
Each Markov chain $\mchain$, induced by a PVASS satisfies the following:
\begin{enumerate}[$\bullet$]
\item $\mchain$ is coarse.
\item $\mchain$ is effective, finitely spanning and globally coarse w.r.t. any upward-closed set of final states.
\end{enumerate}
\end{thm}
This combined with Lemma~\ref{lem:decisive} and Lemma~\ref{lem:gcoarse_decisive}
yields the following corollary.

\begin{cor}
\label{pvass:decisive}
Each Markov chain induced by a PVASS is decisive 
(and thus, by Lemma~\ref{lem:decisive}, strongly decisive) 
w.r.t. any upward-closed set of final states.
\end{cor}

\subsection{Probabilistic Lossy Channel Systems}

\noindent
A {\it Lossy Channel System (LCS)} consists of a finite-state process
operationg on a finite set of channels, each of which behaves as
a FIFO buffer which is unbounded and unreliable in the sense that it can
spontaneously lose messages \cite{AbJo:lossy:IC,CeFiPu:unreliable:IC}.

{\it Probabilistic lossy channel systems (PLCS)} are a generalization of LCS
to a probabilistic model for message loss and choice of transitions. 
There exist several variants of PLCS which differ in
how many messages can be lost, with which probabilities, and in which
situations, and 
whether normal transitions are subject to non-deterministic or
probabilistic choice.
We consider a partial order on channel contents, defined by $w_1 \le w_2$
iff $w_1$ is a (not necessarily continuous) substring of $w_2$.

The most common PLCS model is the one from
\cite{Parosh:Alex:PLCS,Schnoeblen:plcs,Rabinovich:plcs}, where
each message in transit independently has the probability $\lambda>0$ of
being lost in every step, and the transitions are subject to probabilistic
choice in a similar way as for PVASS.\/ However, the definition of PLCS in
\cite{Parosh:Alex:PLCS,Schnoeblen:plcs,Rabinovich:plcs}
assumes that messages can be lost only after discrete steps, but not before
them. Thus, since no messages can be lost before the first discrete
step, the set $\{s\in S:s\models\exists\eventually\fstates\}$ of
predecessors of a given set $\fstates$ of target states is generally not
upward-closed w.r.t.\ $\le$.

Here we assume a more realistic PLCS model where messages can be lost before
and after discrete steps. This PLCS model is also closer to the classic 
non-probabilistic LCS model where also messages can be lost before and after
discrete steps \cite{AbJo:lossy:IC,CeFiPu:unreliable:IC}. So we obtain that
the set $\{s\in S:s\models\exists\eventually\fstates\}$ is always
upward-closed w.r.t.\ $\le$.

\begin{defi}
Formally, a PLCS is a tuple $\plcs=\plcstuple$ where $\lcsstates$ is a
finite set of \emph{control-states}, 
$\channels$ is a finite set of unbounded \emph{fifo-channels}, $\msgs$ is a finite
set called the \emph{message alphabet}, $\lcstransitions$ is a set of
\emph{transitions}, $0<\lossp<1$ is the \emph{message loss rate}, and
$\lcsweight:\lcstransitions\to\nat_{>0}$ is the transition weight function.
Each transition $\lcstransition\in\lcstransitions$ is of the form
$\lcsstate\transitionx{\op}\lcsstate'$, where
$\lcsstate,\lcsstate'\in\lcsstates$ and $\op$ is an operation 
of one of the following froms: 
$\channel!\msg$ (send message $\msg\in\msgs$ in channel
$\channel\in\channels$), $\channel?\msg$ (receive message $\msg$ from
channel $\channel$), or $\nop$ (do not modify the channels).

A PLCS $\plcs=\plcstuple$ induces a transition system $\tssystem=\tstuple$, where
$\states=\lcsstates\times(\msgs^*)^\channels$.
That is, each state in $\states$ consists of a control-state and a function
that assigns a finite word over the message alphabet to each channel called \emph{channel state}.
We define two transition relations $\transition{}_d$ (called `discrete transition')
and $\transition{}_l$ (called `loss transition'), where $\transition{}_d$
models the sending and receiving of
messages and transitions taken in the underlying 
control structure, and $\transition{}_l$ models probabilistic losses of messages.

The relation $\transition{}_d$ is defined as follows.
If $\state=(\lcsstate,\chassignment),
\state'=(\lcsstate',\chassignment')\in\states$, then there is a
transition $\state\transition{}_d\state'$ in the transition system iff one of the
following holds:
\begin{enumerate}[$\bullet$]
\item $\lcsstate\transitionx{\nop}\lcsstate'$ and
  $\chassignment=\chassignment'$;
\item $\lcsstate\transitionx{\channel!\msg}\lcsstate'$,
  $\chassignment'(\channel)=\chassignment(\channel)\msg$, and for all
  $\channel'\in\channels-\{\channel\}$,
  $\chassignment'(\channel')=\chassignment(\channel')$;
\item $\lcsstate\transitionx{\channel?\msg}\lcsstate'$,
  $\chassignment(\channel)=\msg\chassignment'(\channel)$, and for all
  $\channel'\in\channels-\{\channel\}$,
  $\chassignment'(\channel')=\chassignment(\channel')$.
\end{enumerate}
We assume, without loss of generality,
that there are no deadlocks.
This can be guaranteed by adding self-loops
$\lcsstate\transitionx{\nop}\lcsstate$ if necessary.

If several discrete transitions are
enabled at the same configuration then the next transition is chosen
probabilistically. The probability
($\probability_d$) that a
particular transition is taken is given by the weight of this transition,
divided by the sum of the weights of all currently enabled transitions.
Since there are no deadlocks, this is well defined.

The transition $\transition{}_l$ models probabilistic losses of messages.
We extend the subword ordering $\le$
on words first to channel states
$\chassignment,\chassignment':\channels\to\msgs^*$ by
$\chassignment\le\chassignment'$ iff
$\chassignment(\channel)\le\chassignment'(\channel)$ for all
channels $\channel\in\channels$, and then to the transition system states
$\state=(\lcsstate,\chassignment),\state'=(\lcsstate',\chassignment')\in\states$
by $\state\le\state'$ iff $\lcsstate=\lcsstate'$, and
$\chassignment\le\chassignment'$.
For any $\state=(\lcsstate,\chassignment)$ and any $\chassignment'$
such that $\chassignment'\le\chassignment$, there is a transition
$\state\transition{}_l(\lcsstate,\chassignment')$.
The probability of loss transitions is given by
$\probability_l((\lcsstate,\chassignment),(\lcsstate,\chassignment')) =
a\cdot\lossp^b\cdot(1-\lossp)^c$, where $a$ is the number of ways to
obtain $\chassignment'$ by losing messages in $\chassignment$, $b$ is
the total number of messages lost in all channels, and $c$ is the
total number of messages in all channels of $\chassignment'$.

The PLCS induces a Markov chain by alternating the probabilistic transition
relations $\transition{}_l$ and $\transition{}_d$ in such a way that 
message losses can occur before and after every discrete transition, i.e., we
consider transition sequences in $\transition{}_l (\transition{}_d \transition{}_l)^*$.
\end{defi}

We say that a set of target states $\fstates$ is {\em effectively representable}
if a finite set $\fstates'$ can be computed s.t. $\fstates'\uparrow =
\fstates\uparrow$, i.e., their upward-closures are equivalent.
(For instance, any context-free language is effectively representable \cite{Courcelle:obstruction}.)
In \cite{Parosh:Bengt:Karlis:Tsay:general:IC} it is shown that
a Markov chain, induced by a PLCS is effective w.r.t. any effectively 
representable set $\fstates$.

However, many of our results do not strongly depend on a particular PLCS
model. The only crucial aspects are the existence of a finite attractor in the 
induced Markov chain (most PLCS models have it) and the standard 
decidability results of the 
underlying non-probabilistic LCS~\cite{AbJo:lossy:IC,CeFiPu:unreliable:IC}.
In \cite{Parosh:Alex:PLCS}, it is shown that
each Markov chain induced by a PLCS contains a finite attractor.

\begin{thm}
\label{plcs:theorem}
Each Markov chain induced by a PLCS contains a finite attractor and
is effective w.r.t.\ any effectively representable set of global states $\fstates$.
\end{thm}

From this result, Lemma~\ref{lem:decisive} and Lemma~\ref{lem:attractor_decisive}, 
we obtain the following corollary.

\begin{cor}
\label{plcs:decisive}
Each Markov chain induced by a PLCS is decisive w.r.t.\ every set $\fstates$ 
and thus strongly decisive w.r.t.\ every set $\fstates$.
\end{cor}

The PLCS models used here (and in \cite{Parosh:Alex:PLCS,Schnoeblen:plcs,Rabinovich:plcs}) differ from the
more unrealistic models considered previously in \cite{ABIJ:problossy,BaEn:plcs}.
In \cite{BaEn:plcs} at most one message could be lost during any step and in
\cite{ABIJ:problossy} messages could be lost only during send
operations. If one assumes a sufficiently high probability ($> 0.5$) of message loss for
these models then they also contain a finite attractor.
Another different PLCS model was studied in \cite{BerSch-avis2004}. It has the
same kind of probabilistic message loss as our PLCS, but differs in 
having nondeterministic choice (subject to external schedulers) instead 
of probabilistic choice for the transitions, and thus does not yield a Markov
chain, but a Markov decision process. 
Another difference is that the model of \cite{BerSch-avis2004} allows 
(and in some cases requires) idle transitions which are not present in our PLCS model. However, 
for any scheduler, the PLCS model of \cite{BerSch-avis2004} also has a finite
attractor (w.r.t.\ the system-state, though not necessarily w.r.t.\ the state of the scheduler).

\subsection{Noisy Turing Machines}

\noindent
\emph{Noisy Turing Machines (NTM)} were introduced in ~\cite{asarin2005noisy}.
They are \emph{Turing Machines} augmented by an additional parameter $\noise>0$ 
giving the \emph{noise level}.
Each transition of an NTM consists of two steps.
First, in the \emph{noisy step} the tape cells are subjected to noise.
In this manner, each symbol in each tape may change independently and uniformly 
with probability $\noise$ to any other symbol in the tape alphabet (possibly the same 
as before).
Then, in the \emph{normal step}, the NTM proceeds like a normal Turing machine.

\emph{Probabilistic Turing Machines (PTM)} ~\cite{Leeuwprobabilistic}, which are 
Turing machines where transitions are random choices among finitely many 
alternatives, are more general than the model of ~\cite{asarin2005noisy}.
In fact, any NTM can be simulated by a PTM by adding extra steps where the machine 
makes a pass over the tapes changing the symbols randomly.
However, as described below, general PTM do not satisfy our conditions.

\paragraph{\bf Probabilistic NTM} In this paper, we adopt the model of \emph{Probabilistic 
Noisy Turing Machines (PNTM)} which are a generalization of NTM.
In a PNTM, the transitions are similar to those of an NTM except that normal steps are 
subject to probabilistic choices.
Formally, a PNTM $\ntm$ is a tuple $\ntmtuple$ where $\ntmstates$ is a finite set of 
control-states, $\inputalphabet$ is the input alphabet, $\tapealphabet\supseteq\inputalphabet\cup\blank$ 
(where $\blank$ is the blank symbol) is the tape alphabet, $\tapes$ is the number of tapes, 
$\ntmtransitions\subseteq\ntmstates\times\tapealphabet^{\tapes}\times\ntmstates\times\tapealphabet^{\tapes}\times\set{-1,0,1}^{\tapes}$ 
is the transition relation, $\noise$ is the noise 
level and $\ntmweight:\ntmtransitions\to\nat_{>0}$ is the weight function.
The probability of a transition $\ntmtransition\in\ntmtransitions$ is given by 
comparing the weight $\ntmweight(\ntmtransition)$ to the weights of all possible 
alternatives.

Assume a PNTM $\ntm=\ntmtuple$.
A \emph{global state} of $\ntm$ can be represented by a triple: (i) the control-state, 
(ii) the current time, and (iii) an $\tapes$-tuple of \emph{tape configurations}.
A tape configuration is a triple: (i) the head position; (ii) a finite word 
$\omega\in\tapealphabet^*$ representing the content of all cells visited by 
the head so far; and (iii) a $|\omega|$-tuple of natural numbers, each 
representing the last point in time the head visited the corresponding cell.
For a set $\qntmstates\subseteq\ntmstates$, we let $\qntmstates$-states denotes the 
set of all global states whose control-states are in $\qntmstates$.

For a PNTM $\ntm=\ntmtuple$, we use $\graph{\ntm}$ to denote the graph obtained 
from $\ntm$ by abstracting away the memory tapes. 
Formally, $\graph{\ntm}$ is the tuple $\tuple{\ntmstates,\ntmtransitions'}$ where 
$\ntmstates$ is the set of control-states of the underlying PNTM $\ntm$, and 
$\ntmtransitions'\subseteq\ntmstates\times\ntmstates$ is obtained form the 
transition relation of $\ntm$ by projection.
Observe that any path in $\graph{\ntm}$ corresponds to a possible sequence of 
transitions in $\ntm$ since in each step, symbols under the reading heads can 
always change enabling the desired transition. 
Such statements are not possible for general PTM, since the reachability of any 
control-state still depends on the tape configurations and thus cannot be reduced 
to a reachability question in the induced graph.
Nevertheless, for PNTM the following holds.
\begin{lem}
\label{lem:ntm_ctl}
Given a PNTM $\ntm=\ntmtuple$, for any \ctlstar\ formula $\phi$ over sets 
$\fstates_1\cdots\fstates_n$ of $\qntmstates_1\text{-states}\cdots\qntmstates_n\text{-states}$ 
for some $\qntmstates_1\cdots\qntmstates_n\subseteq\ntmstates$, the set of global states 
$\state$ satisfying $\state\models\phi$ is computable.
\end{lem}
\begin{proof}
Observe that checking $\state\models\phi$ is equivalent to checking, in $\graph{\ntm}$, 
$\ntmstate\models\phi'$ where $\ntmstate$ is the control-state in $\state$ and $\phi'$ 
is the formula obtained from $\phi$ by replacing all occurrences of 
$\fstates_1\cdots\fstates_n$ by $\qntmstates_1\cdots\qntmstates_n$ respectively. 
It follows that the set of global states satisfying $\phi$ is exactly the set of 
$\qntmstates$-states such that for any $\ntmstate\in\qntmstates$, $\ntmstate\models\phi'$.
Since $\graph{\ntm}$ is finite, the result follows by decidability of \ctlstar 
model-checking in finite-state systems (\cite{CGP:book}).
\end{proof}

A PNTM $\ntm$ induces a Markov chain $\mchain=\mctuple$ on the set of global states.
Each transition in $\mchain$ is also a combination of a noisy step followed by a 
normal step.
However, in the noisy steps, we assume that cells not under the reading heads 
are not subjected to noise.
Observe that this is different than the way noise is added in the model of 
~\cite{asarin2005noisy} where, for instance, all cells are subject to noise.
Intuitively, the noise doesn't affect the computations of the underlying Turing 
machine unless it changes a cell which is going to be visited by the reading head.
Now, whether the content of that cell changes when the reading head reaches 
it or has changed in the previous steps; the resulting computation is the same.

In order to compensate for the missing noise, we assume a higher noise 
probability for the cell under the head.
If the cell was last visited $k$ time units ago, then we increase the noise 
probability to $1-(1-\noise)^k$.
The probability of a transition in the induced Markov chain is obtained by 
multiplying the noise probability by the probability of the normal step 
described earlier. 

\begin{thm}
\label{ntm:theorem}
Each Markov chain induced by a PNTM $\ntm=\ntmtuple$ is coarse, effective and finitely spanning 
with respect to any set of $\qntmstates$-states for some $\qntmstates\subseteq\ntmstates$.
\end{thm}
\begin{proof}
Assume a PNTM $\ntm=\ntmtuple$, a set $\qntmstates\subseteq\ntmstates$ and the 
induced Markov chain $\mchain$.
Let $\fstates$ be the set of $\qntmstates$-states. 
Effectiveness of $\mchain$ w.r.t. $\fstates$ follows from the definition and Lemma~\ref{lem:ntm_ctl}.
For any state $\state\in\states$, if 
$\state\models\exists\eventually\fstates$ then there is a path in $\graph{\ntm}$ 
from the control-state of $\state$ to a control-state in $\qntmstates$.
Such a path has length at most $N=|\graph{\ntm}|$.
Thus $\mchain$ has span $N$ with respect to $\fstates$.
Along this path, it is possible that, in each step, each symbol under a reading head is subject 
to noise.
Since in each step, $\tapes$ cells are subject to noise and each happens with probability 
$\ge \noise$, it follows that the probability of each successor is 
$\ge (\noise/|\tapealphabet|)^\tapes$.
This gives the coarseness of $\mchain$.
\end{proof}
This, combined with Lemma~\ref{lem:globally_coarse} and Lemma~\ref{lem:gcoarse_decisive}, 
yields the following corollary.

\begin{cor}
\label{ntm:decisive}
Each Markov chain induced by a PNTM $\ntm=\ntmtuple$ is decisive and 
thus (by Lemma~\ref{lem:decisive}) strongly 
decisive with respect to any set of $\qntmstates$-states for some 
$\qntmstates\subseteq\ntmstates$.
\end{cor}

\section{Qualitative Reachability}
\label{sec:quali_reach}

\noindent
We consider the qualitative reachability problem for Markov chains, i.e.,
the problem if a given set of final states is eventually reached with 
probability 1, or probability 0, respectively.

\ignore{
\smallskip
\noindent
\framebox{
\begin{minipage}{0.90\hsize}
\qualreach
\\
{\bf Instance}
A Markov chain $\mchain=\mctuple$,a state $\initstate\in\states$ and a set of
final states $\fstates\subseteq\states$.
\newline
{\bf Task}
Decide if 
$\probms{\mchain}(\initstate\models\eventually\fstates) = 1$
(or $=0$).
\end{minipage}
}
\smallskip
}

\problem{\qualreach}{
\item A Markov chain $\mchain=\mctuple$
\item A state $\initstate\in\states$
\item A set of final states $\fstates\subseteq\states$}{
Decide if 
$\probms{\mchain}(\initstate\models\eventually\fstates) = 1$
(or $=0$).}

We show that, for decisive Markov chains, these qualitative questions 
about the Markov chain can be reduced to
structural properties of the underlying transition graph.
The decidability results for PLCS, PVASS and PNTM are summarized 
in Table \ref{tab:quali_res_1}.

{\small\begin{table}
\begin{minipage}{0.97\columnwidth}
\centering
\begin{tabular}{|p{4.3cm}|p{3cm}|p{3cm}|p{3cm}|p{3cm}} \cline{2-4}
 \multicolumn {1}{l|}{}    & {\bf PLCS} & {\bf PVASS} & {\bf PNTM} \\
\hline
$\probms{\mchain}(\initstate\models\eventually\fstates)=1$
&\raggedright
Decidable when $\fstates$ is effectively representable;

see Theorem~\ref{thm:PLCS_reach_isone}. 
&\raggedright
Decidable when $\fstates$ is defined by
control-states~\footnote{In Theorem~\ref{thm:undecidable_upwardclosed}, we
  prove that this is undecidable when $\fstates$ is a general upward-closed
  set.}\ ;

see Theorem~\ref{thm:0_1_Q_state} 
&
Decidable when $\fstates$ 

is defined by 

control-states;

see Theorem~\ref{thm:ntm_1_reach} \\
\cline{1-4}
$\probms{\mchain}(\initstate\models\eventually\fstates)=0$
&\raggedright
Decidable when $\fstates$ is effectively representable;

see Theorem~\ref{thm:plcs_reach_iszero} 
&\raggedright
Decidable when $\fstates$ is expressible in the logic of \cite{Jancar:TCS};

see Theorem~\ref{thm:pvass_reach_iszero} 
&
Decidable when $\fstates$

is defined by

control-states;

by Theorem~\ref{thm:pntm_reach_iszero} \\
\cline{1-4}
\hline
\end{tabular}
\caption{Decidability results for qualitative reachability.}
\label{tab:quali_res_1}
\end{minipage}
\end{table}
}

First we consider the problem if 
$\probms{\mchain}(\initstate\models\eventually\fstates) = 1$.
The following Lemma holds for any Markov chain and any set of states $\fstates$.

\begin{lem}\label{lem:uFbF}
$\probms{\mchain}(\initstate\models\eventually\fstates) = 1$ 
implies
$\initstate\not\models \unreachable{\fstates}\before\fstates$.
\end{lem}
\begin{proof}
If $\initstate\models \unreachable{\fstates}\before\fstates$ then there is a path $\pi$ of
finite length from $\initstate$ to some state in $\unreachable{\fstates}$ s.t. $\fstates$ is
not visited in $\pi$.
The set of all continuation runs of the form $\pi\pi'$ thus
has a non-zero probability and never visits $\fstates$. Thus  
$\probms{\mchain}(\initstate\models\eventually\fstates) < 1$.
\end{proof}

The reverse implication of Lemma~\ref{lem:uFbF} holds only for Markov chains
which satisfy certain conditions.

\begin{lem}\label{lem:uFbF_reverse}
Given a Markov chain $\mchain$ and a set $\fstates$ such that $\mchain$ is decisive
w.r.t. $\fstates$, then we have that $\initstate\not\models \unreachable{\fstates}\before\fstates$ 
implies $\probms{\mchain}(\initstate\models\eventually\fstates) = 1$.
\end{lem}
\begin{proof}
From $\initstate\not\models\unreachable{\fstates}\before\fstates$ it follows
that $(\initstate\models\eventually\unreachable{\fstates})\subseteq(\initstate\models\eventually\fstates)$. 
Therefore $\probms{\mchain}(\initstate\models\eventually\fstates)=
\probms{\mchain}(\initstate\models\eventually\fstates\vee\eventually\unreachable{\fstates})$.
Since $\mchain$ is decisive w.r.t. $\fstates$, it follows that $1 = \probms{\mchain}(\initstate\models\eventually\fstates\vee\eventually\unreachable{\fstates})=\probms{\mchain}(\initstate\models\eventually\fstates)$.
\end{proof}

Lemma~\ref{lem:uFbF_reverse} does not hold for general Markov chains; see 
Remark~\ref{rem:gambler} in Section~\ref{sec:quali_repeat}.

Now we apply these results to Markov chains derived from PVASS.\/ Interestingly,
decidability depends on whether the target set $\fstates$ is a set of 
$\qvassstates$-states for some $\qvassstates \subseteq \vassstates$ 
or a general upward-closed set.

\begin{thm}\label{thm:0_1_Q_state}
Given a PVASS $\pvasstuple$ and a set of final states $\fstates$ 
which is the set of $\qvassstates$-states for some
$\qvassstates \subseteq \vassstates$. Then the question
$\probms{\mchain}\left(\initstate\models\eventually\fstates\right) = 1$
is decidable.
\end{thm}
\begin{proof}
Since any set $\fstates$ of $\qvassstates$-states is upward-closed, 
we obtain from Corollary~\ref{pvass:decisive} that the Markov chain derived from
our PVASS is decisive w.r.t. such $\fstates$.
Thus, by Lemma~\ref{lem:uFbF} and Lemma~\ref{lem:uFbF_reverse}, we obtain 
$
\probms{\mchain}\left(\initstate\models\eventually\fstates\right) < 1 \iff
\initstate\models \unreachable{\fstates}\before\fstates
$.
To decide the question 
$\initstate\models \unreachable{\fstates}\before\fstates$, we construct a modified PVASS 
$\tuple{\vassstates,\vars,\vasstransitions',\vassweight'}$ 
by removing all outgoing transitions from states $q \in \qvassstates$.
Formally, $\vasstransitions'$ contains all transitions of the form
$\tuple{\vassstate_1,\op,\vassstate_2} \in \vasstransitions$ with
$\vassstate_1\notin \qvassstates$ and 
$\vassweight'(\vasstransition)=\vassweight(\vasstransition)$ for 
$\vasstransition\in\vasstransitions\cap\vasstransitions'$.
Furthermore, to avoid deadlocks, we add to each state
in $\qvassstates$ a self-loop which does not change the values 
of the variables and whose weight is equal to one.
It follows that $\initstate\models \unreachable{\fstates}\before\fstates$
in $\pvasstuple$ 
iff
$\initstate\models\exists\eventually\unreachable{\fstates}$
in $\tuple{\vassstates,\vars,\vasstransitions',\vassweight'}$.

So we obtain that
$\probms{\mchain}\left(\initstate\models\eventually\fstates\right)=1$ in
$\pvasstuple$ iff $\unreachable{\fstates}$ is not reachable in the VASS 
$\tuple{\vassstates,\vars,\vasstransitions'}$.

The condition if $\unreachable{\fstates}$ is reachable in the VASS
$\tuple{\vassstates,\vars,\vasstransitions'}$ can be checked as
follows. Since, $\fstates$ is upward-closed, the set of
predecessors ${\it Pre}^*(\fstates)$ is upward-closed and can be
effectively constructed by Remark~\ref{rem:pvass}. 
Thus the set $\unreachable{\fstates}=\overline{{\it Pre}^*(\fstates)}$ can be 
effectively described by a formula $\Phi$ in the constraint logic
of \cite{Jancar:TCS}. Finally, by Theorem~\ref{thm:jancar}, it is
decidable if there is a reachable state in $\unreachable{\fstates}$
(i.e., satisfying $\Phi$).
\end{proof}

The situation changes if one considers not a set of $\qvassstates$-states as final states
$F$, but rather some general upward-closed set $F$ (described by its finitely
many minimal elements). In this case one cannot
effectively check the condition $\initstate\models \unreachable{\fstates}\before\fstates$.

\begin{thm}\label{thm:undecidable_upwardclosed}
Given a PVASS $\vass=\pvasstuple$ and an upward-closed set 
of final states $\fstates$ (represented by its finitely many minimal
elements), then the question 
$\probms{\mchain}\left(\initstate\models\eventually\fstates\right) = \rho$
is undecidable for any $\rho \in (0,1]$.
\end{thm}

We will need the following definition for the proof.

\begin{defi}\label{def:2cm_pvass}
\rm
We define a PVASS which weakly simulates a Minsky \cite{Minsky:book}
2-counter machine.
Since this construction will be used in several proofs
(Theorem~\ref{thm:undecidable_upwardclosed} and 
Theorem~\ref{thm:not_constructible}), it contains
a parameter $x >0$ which will be instantiated as needed.

Consider a deterministic Minsky 2-counter machine $M$ with a set of control-states
$K$, initial control-state $k_0$, final accepting state $k_{\it acc}$,
two counters $c_1$ and $c_2$ which are initially zero and 
the usual instructions of increment and test-for-zero-decrement.
For technical reasons we require the following conditions on the behavior of
$M$.
\begin{enumerate}[$\bullet$]
\item
Either $M$ terminates in control-state $k_{\it acc}$, or
\item
$M$ does not terminate. In this case we require that in its
infinite run it infinitely often tests a counter for zero
in a configuration where the tested counter contains a non-zero value.
\end{enumerate}
We call a counter machine that satisfies these conditions an IT-2-counter
machine (IT for `infinitely testing'). Any 2-counter machine $M'$ can be
effectively transformed into an equivalent IT-2-counter machine $M$ by the
following operations. After every instruction of $M'$ we add two new
instructions: First increment $c_1$ by 1 (thus it is now certainly nonzero). Then 
test $c_1$ for zero (this test always yields answer `no'), decrement it
by 1 (so it has its original value again), and then continue with the
next instruction of $M'$. So $M$ is infinitely testing and accepts if and
only if $M'$ accepts. Since acceptance is undecidable for 2-counter machines
\cite{Minsky:book}, it follows that acceptance is also undecidable for IT-2-counter machines. 

We construct a PVASS $\vass=\pvasstuple$ that weakly simulates $M$ as follows.
$\vassstates=K \cup \{k^i\,|\, k \in K, i \in \{1,2\}\} \cup \{{\it err}\}$
and $\vars=\{c_1,c_2\}$.
For every instruction $k_1: c_i := c_i+1; \mbox{goto}\ k_2$ we add a 
transition $(k_1, \op, k_2)$ to $T$, where $\op(c_i)=1$ and $\op(c_j)=0$ for
$j\neq i$ and $\vassweight((k_1, \op, k_2)) := 1$.
For every instruction $k_1: \mbox{If}\ c_i=0\ \mbox{then goto}\ k_2\
\mbox{else}\ c_i:=c_i-1;\mbox{goto}\ k_3$ we add the following transitions
to $T$.
\begin{description}
\item[$\alpha$]
$(k_1, \op_1, k_3)$ with $\op_1(c_i)=-1$ and $\op_1(c_j)=0$ for $j\neq i$ and 
$\vassweight((k_1, \op_1, k_3))=1$.
\item[$\beta$]
$(k_1, \op_2, k_2^i)$ with $\op_2(c_j)=0$ for $j=1,2$ and $\vassweight((k_1, \op_2,
k_2^i))=x$ ($x >0$ is a parameter of $\vassweight$).
\item[$\gamma$]
$(k_2^i,\op_a, k_2)$ with $\op_a(c_j)=0$ for $j=1,2$ and $\vassweight((k_2^i, \op_a,
k_2))=1$.
\item[$\delta$]
$(k_2^i,\op_b, {\it err})$ with $\op_b(c_i)=-1$ and $\op_b(c_j)=0$ 
for $j\neq i$ and $\vassweight((k_2^i, \op_b, {\it err}))=1$.
\end{description}
Finally, to avoid deadlocks in $\vass$, 
we add two self-loops $(k_{\it acc},op_l,k_{\it acc})$ 
and $({\it err},op_l,{\it err})$ with 
$\op_l(c_j)=0$ for $j=1,2$ and weight 1.
\end{defi}

\begin{proof} (of Theorem~\ref{thm:undecidable_upwardclosed})
Since $\fstates$ is upward-closed, we obtain from 
Corollary~\ref{pvass:decisive} that the Markov chain derived from our PVASS 
is decisive w.r.t. $\fstates$.
Thus, by Lemma~\ref{lem:uFbF} and Lemma~\ref{lem:uFbF_reverse}, we have 
$\probms{\mchain}\left(\initstate\models\eventually\fstates\right) < 1 \iff
\initstate\models \unreachable{\fstates}\before\fstates$.
Now we show that the condition $\initstate\models \unreachable{\fstates}\before\fstates$ is
undecidable if $\fstates$ is a general upward-closed set. 
We use the IT-2-counter machine $M$ and the PVASS $\vass$ from
Def.~\ref{def:2cm_pvass} and instantiate the parameter $x:=1$.
Let $\fstates$ be the set of configurations where transitions of type
$\delta$ are enabled. This set is upward-closed, because of the monotonicity
of VASS, and effectively constructible (i.e., its finitely many minimal elements). 
It follows directly from the construction in Def.~\ref{def:2cm_pvass}
that a transition of type $\delta$ is enabled if and only if the
PVASS has been unfaithful in the simulation of the 2-counter machine,
i.e., if a counter was non-zero and a `zero' transition 
(of type $\beta$) has wrongly been taken instead of
the correct `decrement' transition (of type $\alpha$).

If the 2-counter machine $M$ accepts then there is a run 
in the PVASS $\vass$ which faithfully simulates the run 
of $M$ and thus never enables transitions of type $\delta$ and thus
avoids the set $\fstates$. Since the $k_{\it acc}$-states have no
outgoing transitions (except for the self-loop),
they are trivially contained in $\unreachable{\fstates}$.
Thus $\initstate\models \unreachable{\fstates}\before\fstates$.

If the 2-counter machine $M$ does not accept then its run is
infinite. By our convention in Def.~\ref{def:2cm_pvass}, $M$ is an 
IT-2-counter machine and every infinite
run must contain infinitely many non-trivial tests for zero. Thus in our PVASS 
$\vass$, the set $\fstates$ is reachable from every reachable state $s'$
which was reached in a faithful simulation of $M$, i.e., without visiting
$\fstates$ before. Therefore in $\vass$ the set $\unreachable{\fstates}$ cannot be
reached unless $\fstates$ is visited first, and so we get
$\initstate \not\models \unreachable{\fstates}\before\fstates$.

We obtain that $M$ accepts iff
$\initstate\models \unreachable{\fstates}\before\fstates$
iff
$\probms{\mchain}\left(\initstate\models\eventually\fstates\right) < 1$.
This proves the undecidability of the problem for the case of $\rho=1$.

To show the undecidability for general $\rho \in (0,1]$ we modify the
construction as follows.
Consider a new PVASS $\vass'$ which with probability $\rho$ does the same as
$\vass$ defined above and
with probability $1-\rho$ immediately goes to the accepting state $k_{\it acc}$.
Then the IT-2-CM $M$ accepts iff 
$\probms{\mchain'}\left(\initstate\models\eventually\fstates\right) \neq \rho$.
\end{proof}

Notice the difference between Theorem~\ref{thm:0_1_Q_state} and
Theorem~\ref{thm:undecidable_upwardclosed} in the case of $\rho=1$. 
Unlike for non-probabilistic VASS, reachability of control-states and
reachability of upward-closed sets cannot be effectively expressed in terms of each other
for PVASS.

\begin{thm}
\label{thm:PLCS_reach_isone}
Consider a PLCS $\plcs$ and an effectively representable set of final states
$\fstates$. Then the 
question $\probms{\mchain}\left(\initstate\models\eventually\fstates\right) = 1$
is decidable.
\end{thm}
\begin{proof}
By Corollary~\ref{plcs:decisive}, the Markov chain induced by $\plcs$ is decisive w.r.t. such $\fstates$. 
Thus we obtain from Lemma~\ref{lem:uFbF} and Lemma~\ref{lem:uFbF_reverse}
that $\probms{\mchain}\left(\initstate\models\eventually\fstates\right) = 1$
iff $\initstate\not\models \unreachable{\fstates}\before\fstates$. 
This condition can be checked with a standard construction for LCS
(from \cite{Parosh:Alex:PLCS})
as follows. First one can effectively compute the set
$\unreachable{\fstates} = \overline{{\it Pre}^*(\fstates)}$ using the techniques
from, e.g., \cite{AbJo:lossy:IC}. Next one computes the 
set $X$ of all configurations from which it is possible to reach 
$\unreachable{\fstates}$ without passing through $\fstates$. 
This is done as follows. Let $X_0 := \unreachable{\fstates}$ and 
$X_{i+1} := X_i\!\uparrow\,\cup\, ({\it Pre}(X_i) \cap \overline{F})\!\uparrow$.
Since all $X_i$ are upward-closed, this construction converges at some finite 
index $n$, by Higman's Lemma \cite{Higman:divisibility}.
We get that $X = X_n$ is effectively constructible.
Finally we have that
$\probms{\mchain}\left(\initstate\models\eventually\fstates\right) = 1$
iff $\initstate \notin X$, which can be effectively checked. 
\end{proof}

Notice that, unlike in earlier work \cite{Parosh:Alex:PLCS,Schnoeblen:plcs},
it is not necessary to compute the finite attractor of the PLCS-induced Markov chain
for Theorem~\ref{thm:PLCS_reach_isone}. It suffices to know that
it exists. For PLCS it is very easy to construct the finite attractor,
but this need not hold for other classes of systems with attractors.
However, the criterion given by Lemma~\ref{lem:uFbF} and
Lemma~\ref{lem:uFbF_reverse} always holds.

\begin{thm}
\label{thm:ntm_1_reach}
For a PNTM $\ntm=\ntmtuple$, the question $\probms{\mchain}\left(\initstate\models\eventually\fstates\right) = 1$ 
is decidable for any set $\fstates$ of $\qntmstates$-states for some $\qntmstates\subseteq\ntmstates$.
\end{thm}
\begin{proof}
By Corollary~\ref{ntm:decisive}, we obtain that the Markov chain $\mchain$ derived from $\ntm$ is 
decisive w.r.t. $\fstates$.
This combined with Lemma~\ref{lem:uFbF} and Lemma~\ref{lem:uFbF_reverse} yields $
\probms{\mchain}\left(\initstate\models\eventually\fstates\right) < 1 \iff
\initstate\models \unreachable{\fstates}\before\fstates
$.
Observe that since $\fstates$ is a set of $\qntmstates$-states, we obtain by Lemma~\ref{lem:ntm_ctl} that we can 
compute a set $\qntmstates'\subseteq\ntmstates$ such that $\unreachable{\fstates}=\qntmstates'\text{-states}$.
Since $\fstates$ and $\unreachable{\fstates}$ are sets of $\qntmstates$-states and $\qntmstates'$-states respectively, it follows 
by Lemma~\ref{lem:ntm_ctl} that the question $\initstate\models\unreachable{\fstates}\before\fstates$ is decidable.
This gives the result.
\end{proof}

Now we consider the question 
$\probms{\mchain}\left(\initstate\models\eventually\fstates\right) = 0$.
The following property trivially holds for all Markov chains.

\begin{lem}\label{lem:nonzero_iff_reachable}
$\probms{\mchain}\left(\initstate\models\eventually\fstates\right) = 0$ iff 
$\initstate \not\models \exists\eventually\fstates$.
\end{lem}

The reachability problem for Petri nets/VASS 
is decidable \cite{Mayr:SIAM84}, and the following result is a
direct consequence of Lemma~\ref{lem:nonzero_iff_reachable}
and Theorem~\ref{thm:jancar}.

\begin{thm}\label{thm:pvass_reach_iszero}
Given a PVASS $\vass=\pvasstuple$ and a set of final states $\fstates$ 
which is expressible in the constraint logic of \cite{Jancar:TCS}
(in particular any upward-closed set, any finite set, and their complements), then the question
$\probms{\mchain}\left(\initstate\models\eventually\fstates\right) = 0$
is decidable.
\end{thm}

From Lemma~\ref{lem:nonzero_iff_reachable} and the result
that for LCS the set of all predecessors of any effectively representable set 
can be effectively constructed (e.g., \cite{AbJo:lossy:IC}), we get the following.

\begin{thm}\label{thm:plcs_reach_iszero}
Given a PLCS $\plcs$ and a set of final states $\fstates$ which is effectively representable,
then the question
$\probms{\mchain}\left(\initstate\models\eventually\fstates\right) = 0$
is decidable.
\end{thm}

By Lemma~\ref{lem:ntm_ctl}, we obtain the following.

\begin{thm}\label{thm:pntm_reach_iszero}
Given a PNTM $\ntm=\ntmtuple$ and a set $\fstates$ of $\qntmstates$-states for some 
$\qntmstates\subseteq\ntmstates$, then the question 
$\probms{\mchain}\left(\initstate\models\eventually\fstates\right) = 0$
is decidable.
\end{thm}

\section{Qualitative Repeated Reachability}
\label{sec:quali_repeat}
\noindent
Here we consider the qualitative repeated reachability problem 
for Markov chains, i.e.,
the problem if a given set of final states $\fstates$ is visited infinitely
often with probability 1, or probability 0, respectively.

\ignore{
\smallskip
\noindent
\framebox{
\begin{minipage}{0.90\hsize}
\qualrepreach
\\
{\bf Instance}
A Markov chain $\mchain=\mctuple$, 
a state $\initstate\in\states$ and a set of final states $\fstates\subseteq\states$.
\newline
{\bf Task}
Decide if 
$\probms{\mchain}\left(\initstate\models \always\eventually\fstates\right) = 1$
(or $=0$).
\end{minipage}
}
\smallskip
}

\problem{\qualrepreach}{
\item A Markov chain $\mchain=\mctuple$
\item A state $\initstate\in\states$
\item A set of final states $\fstates\subseteq\states$}{
Decide if 
$\probms{\mchain}\left(\initstate\models \always\eventually\fstates\right) = 1$
(or $=0$).}

We show that, for decisive Markov chains, these qualitative questions 
about the Markov chain can be reduced to
structural properties of the underlying transition graph.
The decidability results for PLCS, PVASS and PNTM are summarized 
in Table \ref{tab:quali_res_2}.

{small\begin{table}
\begin{minipage}{0.98\columnwidth}
\centering

\begin{tabular}{|p{4.3cm}|p{3cm}|p{3cm}|p{3cm}|p{3cm}} \cline{2-4}
 \multicolumn {1}{l|}{}    & {\bf PLCS} & {\bf PVASS} & {\bf PNTM} \\
\hline
$\probms{\mchain}(\initstate\models\always\eventually\fstates)=1$
&\raggedright
Decidable when $\fstates$ is effectively representable;

see Theorem~\ref{thm:PLCS_repeat_isone} 
&\raggedright
Decidable when $\fstates$ is upward- closed;

see Theorem~\ref{thm:0_1_upwardclosed_repeat} 
&
Decidable when

$\fstates$ is defined by

control-states;

see Theorem~\ref{thm:PNTM_repeat_isone} \\
\cline{1-4}
$\probms{\mchain}(\initstate\models\always\eventually\fstates)=0$
&\raggedright
Decidable when $\fstates$ is effectively representable

see Theorem~\ref{thm:PLCS_repeat_iszero} 
&
Open problem
&
Decidable when

$\fstates$ is defined by 

control-states;

see Theorem~\ref{thm:PNTM_repeat_iszero} \\
\cline{1-4}
\hline
\end{tabular}
\caption{Decidability results for qualitative problems of repeated reachability.}
\label{tab:quali_res_2}
\end{minipage}
\end{table}
}

First we consider the problem if 
$\probms{\mchain}\left(\initstate\models\always\eventually\fstates\right) = 1$.
The following lemma holds for any Markov chain and any set of states $\fstates$.

\begin{lem}\label{lem:always_again}
Let $\mchain=\mctuple$ be a Markov chain and $\fstates \subseteq \states$.
Then 
$\probms{\mchain}\left(\initstate\models\always\eventually\fstates\right) = 1$ 
implies
$\initstate\models \forall\always \exists\eventually \fstates$.
\end{lem}
\begin{proof}
Suppose that $\initstate\not\models \forall\always \exists\eventually \fstates$.
Then $\initstate\models \exists\eventually\forall\always \neg\fstates$.
Thus there exists a finite path $\pi$ starting from $\initstate$ leading to a 
state $\state$ s.t. $\state \models \forall\always \neg\fstates$.
The set of all $\initstate$-runs of the form $\pi\pi'$ (for any $\pi'$) has non-zero probability
and they all satisfy $\neg\always\eventually\fstates$. 
So we get that
$\probms{\mchain}\left(\initstate\models\always\eventually\fstates\right) < 1$.
\end{proof}

The reverse implication of Lemma~\ref{lem:always_again} does not hold
generally, but it is true for strongly decisive Markov chains.

\begin{lem}\label{lem:always_again_reverse}
Given a Markov chain $\mchain$ and a set $\fstates$ 
such that $\mchain$ is strongly decisive w.r.t.\ $\fstates$,
 then we have that $\initstate\models \forall\always \exists\eventually \fstates$
implies $\probms{\mchain}\left(\initstate\models\always\eventually\fstates\right) = 1$.
\end{lem}
\begin{proof}
We show that
$\probms{\mchain}(\initstate\models\eventually\always\neg\fstates)=0$ 
which implies the result.

If $\initstate\models \forall\always \exists\eventually \fstates$
then every state $\state$ reached by runs in
$(\initstate\models\eventually\always\neg\fstates)$ satisfies
$\state\models\exists\eventually\fstates$.
Therefore $(\initstate\models\eventually\always\neg\fstates)
\subseteq (\initstate\models\always\neg\unreachable{\fstates})$. 

Since the Markov chain is strongly decisive, 
$\probms{\mchain}(\initstate\models\eventually\unreachable{\fstates}\vee\always\eventually\fstates)=1$
and thus
$\probms{\mchain}(\initstate\models\always\neg\unreachable{\fstates}\wedge\eventually\always\neg
\fstates)=0$.
It follows from the inclusion above that
$0 = \probms{\mchain}(\initstate\models\always\neg\unreachable{\fstates}\wedge\eventually\always\neg
\fstates) =
\probms{\mchain}(\initstate\models\always\neg\unreachable{\fstates})
\ge 
\probms{\mchain}(\initstate\models\eventually\always\neg\fstates)$. 
\end{proof}

\begin{rem}\label{rem:gambler}
Neither Lemma~\ref{lem:uFbF_reverse} nor Lemma~\ref{lem:always_again_reverse}
hold for general Markov chains. 
A counterexample is the Markov chain $\mchain=\mctuple$ of the `gambler's
ruin' problem \cite{Feller:book}
where $\states=\nat$, $\probability(i,i+1):=x$, $\probability(i,i-1):=1-x$ for $i \ge 1$ and
$\probability(0,0)=1$ and
$\fstates:=\{0\}$, for some parameter $x \in [0,1]$. It follows 
that $\unreachable{\fstates}=\emptyset$ for $x<1$.
If $x \in [0,1/2]$ then 
$\probms{\mchain}\left(1 \models \eventually\fstates\right)=1$
and for $x > 1/2$ one has 
$\probms{\mchain}\left(1 \models \eventually\fstates\right)=(1-x)/x$.

For $x \in (1/2,1)$ one has that $1 \not\models \unreachable{\fstates}\before\fstates$,
but $\probms{\mchain}\left(1 \models \eventually\fstates\right) = (1-x)/x < 1$.
Similarly, although $1 \models \forall\always \exists\eventually \fstates$,
one still has 
$\probms{\mchain}\left(1 \models\always\eventually\fstates\right)
\le \probms{\mchain}\left(1 \models \eventually\fstates\right) = (1-x)/x < 1$.
\end{rem}

Now we show that it is decidable if a PVASS almost certainly reaches an
upward-closed set of final states infinitely often. 

\begin{thm}\label{thm:0_1_upwardclosed_repeat}
Let $\vass=\pvasstuple$ be a PVASS and $\fstates$ an upward-closed set of
final states.
Then the question
$\probms{\mchain}\left(\initstate\models\always\eventually\fstates\right) = 1$
is decidable.
\end{thm}
\begin{proof}
Since $\fstates$ is upward-closed, we obtain from Corollary~\ref{pvass:decisive}
that the Markov chain derived from our PVASS is strongly decisive w.r.t.\ $\fstates$. Thus it
 follows from Lemma~\ref{lem:always_again} and Lemma~\ref{lem:always_again_reverse} that
$
\probms{\mchain}\left(\initstate\models\always\eventually\fstates\right) = 1
\iff
\initstate\models \forall\always \exists\eventually \fstates
$.
This condition can be checked as follows. Since $\fstates$ is upward-closed 
and represented by its finitely many 
minimal elements, the set ${\it Pre}^*(\fstates)$ is upward-closed and
effectively constructible by Remark~\ref{rem:pvass}.
Then $\unreachable{\fstates} =\overline{{\it Pre}^*(\fstates)}$
is downward-closed and effectively representable by a formula $\Phi$ in the 
constraint logic of \cite{Jancar:TCS}. We get that
$\initstate\models \forall\always \exists\eventually \fstates$
iff $\initstate \not\models \exists\eventually \unreachable{\fstates}$,
i.e., if there is no reachable state that satisfies $\Phi$.
This is decidable by Theorem~\ref{thm:jancar}.
\end{proof}

Notice the surprising contrast of the decidability of repeated reachability
of Theorem~\ref{thm:0_1_upwardclosed_repeat} to the undecidability
of simple reachability in Theorem~\ref{thm:undecidable_upwardclosed}.

Now we show the decidability result for PLCS.

\begin{thm}\label{thm:PLCS_repeat_isone}
Consider a PLCS $\plcs$ and an effectively representable set of final states $\fstates$.
Then the question
$\probms{\mchain}\left(\initstate\models\always\eventually\fstates\right) = 1$
is decidable.
\end{thm}
\begin{proof}
By Corollary~\ref{plcs:decisive}, $\plcs$ induces a strongly decisive Markov 
chain. Thus, we obtain from 
Lemma~\ref{lem:always_again} and Lemma~\ref{lem:always_again_reverse}
that 
$
\probms{\mchain}\left(\initstate\models\always\eventually\fstates\right) = 1
\iff
\initstate\models \forall\always \exists\eventually \fstates
$.
This condition can be checked as follows.
First one can effectively compute the set
$\unreachable{\fstates} = \overline{{\it Pre}^*(\fstates)}$.
Next one computes the set $X$ of all configurations from 
which it is possible to reach $\unreachable{\fstates}$, i.e., 
$X:={\it Pre}^*(\unreachable{\fstates})$. 
Finally we have that
$\probms{\mchain}\left(\initstate\models\eventually\fstates\right) = 1$
iff 
$\initstate\models \forall\always \exists\eventually \fstates$
iff
$\initstate \notin X$, which can be effectively checked. 
\end{proof}

Similarly as in Theorem~\ref{thm:PLCS_reach_isone},
it is not necessary to compute the finite attractor of the PLCS-induced Markov chain
for Theorem~\ref{thm:PLCS_repeat_isone}.

Next we prove the decidability result for PNTM.

\begin{thm}\label{thm:PNTM_repeat_isone}
Let $\ntm=\ntmtuple$ be a PNTM and $\fstates$ a set of $\qntmstates$-states for 
some $\qntmstates\subseteq\ntmstates$.
Then the question 
$\probms{\mchain}\left(\initstate\models\always\eventually\fstates\right) = 1$
is decidable.
\end{thm}
\begin{proof}
Since $\fstates$ is a set of $\qntmstates$-states, we obtain from Corollary~\ref{ntm:decisive}
that the Markov chain derived from $\ntm$ is strongly decisive w.r.t.\ $\fstates$. Thus it
 follows from Lemma~\ref{lem:always_again} and Lemma~\ref{lem:always_again_reverse} that
$
\probms{\mchain}\left(\initstate\models\always\eventually\fstates\right) = 1
\iff
\initstate\models \forall\always \exists\eventually \fstates
$.
We can check this condition by Lemma~\ref{lem:ntm_ctl}.
\end{proof}

Now we consider the question 
$\probms{\mchain}(\initstate\models\always\eventually\fstates) = 0$.
We start by establishing some connections between 
the probabilities of reaching certain sets at least once or infinitely often.
From the definitions we get the following.

\begin{lem}\label{lem:AA-AR}
For any Markov chain and any set of states $\fstates$, we have
$\probms{\mchain}(\initstate\models\always\eventually\fstates) \neq
0$
implies
$\probms{\mchain}(\initstate\models\eventually\unreachable{\fstates}) 
\neq 1$
\end{lem}

The following lemma implies that the reverse implication 
holds for strongly decisive Markov chains.

\begin{lem}\label{lem:AA-AR_reverse}
Given a Markov chain $\mchain$ which is strongly decisive w.r.t. a given set $\fstates$, then we have that
$
\probms{\mchain}(\initstate\models\always\eventually\fstates) 
= 
1 - \probms{\mchain}(\initstate\models\eventually\unreachable{\fstates})
$.
\end{lem}
\begin{proof}
$1 - \probms{\mchain}(\initstate\models\eventually\unreachable{\fstates})=
\probms{\mchain}(\initstate\models\always\neg\unreachable{\fstates})
=\probms{\mchain}(\initstate\models\always\neg\unreachable{\fstates}\wedge\always\eventually\fstates) 
+ \probms{\mchain}(\initstate\models\always\neg\unreachable{\fstates}\wedge\eventually\always\neg\fstates)$. Since 
the Markov chain is strongly decisive, it follows that 
$\probms{\mchain}(\initstate\models\always\neg\unreachable{\fstates}\wedge\eventually\always\neg\fstates)=0$. Moreover,
 we have 
$(\initstate\models\always\eventually\fstates)\subseteq
 (\initstate\models\always\neg\unreachable{\fstates})$. 
Therefore, $1 - \probms{\mchain}(\initstate\models\eventually\unreachable{\fstates})=\probms{\mchain}(\initstate\models\always\neg\unreachable{\fstates})=\probms{\mchain}(\initstate\models\always\eventually\fstates)$. 
\end{proof}

There is also a correspondence of the condition
$\probms{\mchain}(\initstate\models\always\eventually\fstates) \neq 0$ to a
property of the underlying transition graph.

\begin{lem}\label{lem:EAE_F}
For any Markov chain $\mchain$ and any set of states $\fstates$,  
$\initstate\models\exists\eventually\unreachable{\unreachable{\fstates}}$
implies
$\probms{\mchain}(\initstate\models\eventually\unreachable{\fstates}) 
\neq 1$.
\end{lem}
\begin{proof}
If $\initstate\models\exists\eventually\unreachable{\unreachable{\fstates}}$
then there exists a finite path starting from $\initstate$ and leading to some
state in $\unreachable{\unreachable{\fstates}}$. 
Therefore we obtain $\probms{\mchain}(\initstate\models\eventually
\unreachable{\unreachable{\fstates}})>0$.
Observe that any state $\state$ reached by runs in $(\initstate\models\eventually
\unreachable{\unreachable{\fstates}})$ {\em before} reaching 
$\unreachable{\unreachable{\fstates}}$ satisfies 
$\state\models\exists\eventually\unreachable{\unreachable{\fstates}}$.
So we have $\state\models\exists\eventually\forall\always\exists
\eventually\fstates$ and therefore
$\state\models\exists\eventually\fstates$ and thus
$\state\models\neg\unreachable{\fstates}$.
Furthermore, every state $\state$ reached by runs in $(\initstate\models\eventually
\unreachable{\unreachable{\fstates}})$ {\em after} reaching 
$\unreachable{\unreachable{\fstates}}$ also satisfies
$\state\models\neg\unreachable{\fstates}$,
because, by definition, $\unreachable{\fstates}$ cannot be reached from 
$\unreachable{\unreachable{\fstates}}$.

This yields $(\initstate\models\eventually\unreachable{\unreachable{\fstates}})\subseteq
(\initstate\models\always\neg\unreachable{\fstates})$ which implies that $0<\probms{
\mchain}(\initstate\models\eventually\unreachable{\unreachable{\fstates}})\le
\probms{\mchain}(\initstate\models\always\neg\unreachable{\fstates})$, where the 
first inequality follows from the argument mentioned above.
Finally, we obtain $\probms{\mchain}(\initstate\models\eventually\unreachable{\fstates})
=1-\probms{\mchain}(\initstate\models\always\neg\unreachable{\fstates})
\le 1-\probms{\mchain}(\initstate\models\eventually\unreachable{\unreachable{\fstates}}) < 1$.
\ignore{If $\initstate\models\exists\eventually\unreachable{\unreachable{\fstates}}$
then there exists a finite path $\pi$ starting from $\initstate$ and leading to some
state $\state \in \unreachable{\unreachable{\fstates}}$ and thus $\state \models
\forall\always\exists\eventually\fstates$. Therefore 
$\state\not\models\exists\eventually\unreachable{\fstates}$. Consider the set $R$ of $\initstate$-runs
of the form $\pi\pi'$ for any $\pi'$. This set has a non-zero measure
and all runs in it satisfy $\neg\eventually\unreachable{\fstates}$.  
Thus $\probms{\mchain}(\initstate\models\eventually\unreachable{\fstates}) \le
1-\probms{\mchain}(R) < 1$.}
\end{proof}

\ignore{
The reverse implication of Lemma~\ref{lem:EAE_F} holds only for Markov chains
with a finite attractor, but not generally for globally coarse Markov
chains. This is because global coarseness depends on the set of final states.
Global coarseness of a Markov chain $\mctuple$ does not imply global coarseness 
of $\tuple{\states,\probability,\unreachable{\fstates}}$.
}

If $\mchain$ is decisive w.r.t. $\unreachable{\fstates}$ then the reverse
implication holds.

\begin{lem}\label{lem:EAE_F_reverse}
Given a Markov chain $\mchain$ and a set of states $\fstates$ s.t. $\mchain$ 
is decisive w.r.t. $\unreachable{\fstates}$, then the condition
$\probms{\mchain}(\initstate\models\eventually\unreachable{\fstates}) \neq 1$
implies $\initstate\models\exists\eventually\unreachable{\unreachable{\fstates}}$.
\end{lem}
\begin{proof}
As $\mchain$ is decisive w.r.t.\ $\unreachable{\fstates}$, it follows that
$\probms{\mchain}(\initstate\models\eventually\unreachable{\fstates}\vee\eventually\unreachable{\unreachable{\fstates}})=1$.
Since by assumption $\probms{\mchain}(\initstate\models\eventually\unreachable{\fstates})\neq 1$, 
it follows that
$\probms{\mchain}(\initstate\models\eventually\unreachable{\unreachable{\fstates}})>0$,
which implies $\initstate\models\exists\eventually\unreachable{\unreachable{\fstates}}$.
\end{proof}

\ignore{
Notice that the finite attractor of the Markov chain need not be known or 
constructible. It suffices to know that a finite attractor exists.}

\begin{thm}\label{thm:rep_reach_iszero}
For any Markov chain $\mchain$ and any set $\fstates$ s.t. $\mchain$ is decisive w.r.t.\ $\fstates$
\[
\initstate\models\exists\eventually\unreachable{\unreachable{\fstates}}
\ \Rightarrow
\ \probms{\mchain}(\initstate\models\eventually\unreachable{\fstates}) \neq 1
\iff 
\probms{\mchain}(\initstate\models\always\eventually\fstates) \neq 0
\]
For any Markov chain $\mchain$ and any set $\fstates$ s.t. $\mchain$ is decisive w.r.t.\ $\fstates$
and w.r.t.\ $\unreachable{\fstates}$
\[
\initstate\models\exists\eventually\unreachable{\unreachable{\fstates}}
\iff
\probms{\mchain}(\initstate\models\eventually\unreachable{\fstates}) \neq 1
\iff 
\probms{\mchain}(\initstate\models\always\eventually\fstates) \neq 0
\]
\end{thm}
\begin{proof}
Directly from Lemma~\ref{lem:AA-AR}, \ref{lem:AA-AR_reverse},
\ref{lem:EAE_F} and \ref{lem:EAE_F_reverse}. 
\end{proof}

\begin{rem}
Observe that decisiveness w.r.t.\ a given set $\fstates$ does not imply 
decisiveness w.r.t.\ $\unreachable{\fstates}$. Therefore the reverse implication 
of Lemma~\ref{lem:EAE_F} does not hold in general. In particular, it holds 
for Markov chains with a finite attractor (since they are decisive
w.r.t. every set), but 
not generally when we have global coarseness. This is because global coarseness 
depends on the set of final states. Global coarseness of a Markov chain w.r.t. 
a certain set $\fstates$ does not imply global coarseness w.r.t. the set 
$\unreachable{\fstates}$.
\end{rem}

Now we show the decidability results for PLCS and PNTM.

\begin{thm}\label{thm:PLCS_repeat_iszero}
Consider a PLCS $\plcs$ and an effectively representable set of final states
$\fstates$. Then the question
$\probms{\mchain}\left(\initstate\models\always\eventually\fstates\right) = 0$
is decidable.
\end{thm}
\begin{proof}
By Corollary~\ref{plcs:decisive},
$\plcs$ induces a strongly decisive Markov chain w.r.t.\ every 
set of states. In particular it is decisive 
w.r.t.\ $\fstates$ and $\unreachable{\fstates}$. 
Therefore, by Theorem~\ref{thm:rep_reach_iszero},
it suffices to check if
$\initstate \not\models
\exists\eventually\unreachable{\unreachable{\fstates}}$.
Since the upward-closure of $\fstates$ is effectively constructible,
one can effectively construct a symbolic representation of the set of all states
which satisfy $\exists\eventually\unreachable{\unreachable{\fstates}}$
(using the techniques from, e.g., \cite{AbJo:lossy:IC}) and check if 
$\initstate$ is not in this set. 
\end{proof}

\begin{thm}\label{thm:PNTM_repeat_iszero}
Given a PNTM $\ntm=\ntmtuple$ and a set $\fstates$ of $\qntmstates$-states for some 
$\qntmstates\subseteq\ntmstates$, the question 
$\probms{\mchain}\left(\initstate\models\always\eventually\fstates\right) = 0$
is decidable.
\end{thm}
\begin{proof}
Observe that since $\fstates$ is a set of $\qntmstates$-states, by 
Lemma~\ref{lem:ntm_ctl}, we can construct the set $\qntmstates'\subseteq\ntmstates$ 
such that $\unreachable{\fstates}$ is exactly the set of $\qntmstates'$-states.
By Corollary~\ref{ntm:decisive}, the Markov chain induced by $\ntm$ is 
decisive w.r.t. any set of $\qntmstates''$-states for some $\qntmstates''\subseteq\ntmstates$. 
In particular, it is decisive w.r.t. $\qntmstates'\text{-states}=\unreachable{\fstates}$.
By Theorem~\ref{thm:rep_reach_iszero}, it suffices to check if $\initstate \not\models
\exists\eventually\unreachable{\unreachable{\fstates}}$ which is again decidable by
Lemma~\ref{lem:ntm_ctl}.
\end{proof}

\begin{rem}\label{rem:dec_F_Ftilde}
For PVASS, decidability of
$\probms{\mchain}\left(\initstate\models\always\eventually\fstates\right) = 0$
and the equivalent question 
$\probms{\mchain}(\initstate\models\eventually\unreachable{\fstates}) = 1$ is open.
For an upward-closed set $\fstates$ the set $\unreachable{\fstates}$ is
downward-closed, but in general not a set of $\qvassstates$-states, and thus 
Theorem~\ref{thm:0_1_Q_state} does not always apply.
The question $\probms{\mchain}(\initstate\models\eventually\unreachable{\fstates}) = 1$
can certainly not be reduced to purely structural questions about the
underlying transition system (unlike for PLCS), because it depends on the
exact values of the probabilities, i.e., on the transition weights. 

Consider a PVASS model of the gambler's ruin problem (see
Remark~\ref{rem:gambler}), but let now $\fstates:=\{1,2,\dots\}$ (upward-closed) and thus 
$\unreachable{\fstates}=\{0\}$. 
We have
$\probms{\mchain}(1 \models\eventually\unreachable{\fstates}) = (1-x)/x < 1$ for
$x>1/2$ and $\probms{\mchain}(1 \models\eventually\unreachable{\fstates}) =1$ otherwise.

In particular, for $x > 1/2$, this system is decisive w.r.t.\ $\fstates$, but not
decisive w.r.t.\ $\unreachable{\fstates}$, because
$\unreachable{\unreachable{\fstates}}=\emptyset$
and $\probms{\mchain}(1 \models\eventually\unreachable{\fstates} \vee
\eventually\unreachable{\unreachable{\fstates}})
= \probms{\mchain}(1 \models\eventually\unreachable{\fstates}) < 1$.

Furthermore, for PVASS, the probability 
$\probms{\mchain}\left(\initstate\models\always\eventually\fstates\right)$
cannot be effectively expressed in the first-order theory
of the reals $(\R,+,*,\le)$, as shown in Section~\ref{sec:exact_quant_reach},
Remark~\ref{rem:rep_reach_nonexp}.
\end{rem}

\section{Approximate Quantitative Reachability}
\label{sec:quant_reach}
\noindent
In this section we consider the approximate
quantitative reachability problem.
\ignore{
\vspace{1mm}
\noindent
\framebox{
\begin{minipage}{0.90\hsize}
\aqr
\\
{\bf Instance}
A Markov chain $\mchain=\mctuple$, a state $\initstate\in\states$, a set of states $\fstates\subseteq\states$
and a rational $\error > 0$.
\\
{\bf Task}
Compute a rational $\probapprox$ such that
$\probapprox\leq\probms{\mchain}
\left(\initstate\models\eventually\fstates\right)\leq \probapprox+\error$.
\end{minipage}
}
\vspace{2mm}
}

\problem{\aqr}{
\item A Markov chain $\mchain=\mctuple$
\item A state $\initstate\in\states$
\item A set of states $\fstates\subseteq\states$
\item A rational $\error>0$}{
Compute a rational $\probapprox$ such that
$\probapprox\leq\probms{\mchain}
\left(\initstate\models\eventually\fstates\right)\leq \probapprox+\error$.}

We show that this problem is effectively solvable for PLCS, PVASS and PNTM,
provided that the induced Markov chain is decisive w.r.t.\ $\fstates$.

First, we present a path enumeration
algorithm, based on \cite{IyeNar97},
for solving the problem, and 
then we show that the algorithm is guaranteed to terminate 
for all instances where $\mchain$ is decisive w.r.t. $\fstates$
\footnote{
Termination of the algorithm is stated
(without giving a proof)
in \cite{Rabinovich:plcs}
for the case of Markov chains which contain a finite attractor.}.

Given an effective Markov chain $\mchain=\mctuple$,
a state $\initstate\in\states$, a set $\fstates$ and a positive $\error\in\rat_{>0}$,
the algorithm constructs (a prefix of) the 
reachability-tree, from $\initstate$,
in a breadth-first fashion.
The nodes of the tree are labeled with pairs
$\tuple{\state,r}$ where $\state\in\states$ and
$r$ is the probability of traversing the path from the
root to the current node.
Every node in the tree is labeled with a probability.
This probability is the product of the probabilities of all the transitions
in the path from the root to the node.
The algorithm maintains two variables $\yes$ and $\no$ which
accumulate the probabilities by which the set $\fstates$ 
is certainly reached (and certainly not reached, respectively).
Each step of the algorithm can be implemented 
due to the effectiveness of $\mchain$.
The algorithm runs until we reach a point where the
sum of $\yes$ and $\no$ exceeds $1-\error$.

\algbox{\aqr\label{alg:approx_quant_reach}}
{
{\bf Input}

A Markov chain $\mchain=\mctuple$, a state $\initstate\in\states$,
a set $\fstates\subseteq\states$ and a positive $\error\in\rat_{>0}$ 
such that $\mchain$ is effective w.r.t.\ $\fstates$.

{\bf Return value}

A rational $\probapprox$ such that
$\probapprox\leq\probms{\mchain}
\left(\initstate\models\eventually\fstates\right)\leq \probapprox+\error$

{\bf Variables}

\ind%
$\yes$, $\no$: $\rat$
\hspace{3cm}(initially all are set to 0)

\ind%
${\it store}$: queue with elements in $\states\times\rat$}{
\algline{1.}%
${\it store}:= \tuple{\initstate,1}$

\algline{2.}%
{\bf repeat}

\algline{3.}\ind%
remove $\tuple{\state,r}$ from ${\it store}$

\algline{4.}\ind%
{\bf if} $\state\in\fstates$ {\bf then} $\yes:=\yes+r$

\algline{5.}\ind%
{\bf else if} $\state\in\unreachable{\fstates}$ {\bf then} $\no:=\no+r$

\algline{6.}\ind%
{\bf else} \hspace{1mm}  {\bf for each} $\state'\in\post(\state)$

\algline{7.}\ind\ind%
add $\tuple{\state',r\cdot\probability(\state,\state')}$ 
to the end of ${\it store}$

\algline{8.}%
{\bf until}
$\yes+\no\geq 1-\error$

\algline{9.}%
{\bf return}
$\yes$
}

\smallskip\noindent
We require that the Markov chain is effective w.r.t.\ $\fstates$
so that the condition $\state\in\unreachable{\fstates}$ in line 
5.\ can be effectively checked.

Let $\yes^j(\mchain,\initstate)$ denote the value of variable $\yes$ 
after the algorithm has explored the reachability-tree 
with root $\initstate$ up to depth
$j$ (i.e., any element $\tuple{\state,r}$ in {\it store} is such that
$\initstate\transition{\le j+1}\state$).
We define $\no^j(\mchain,\initstate)$ in a similar manner.
First we show partial correctness of Algorithm~\ref{alg:approx_quant_reach}.

\begin{lem}
\label{quantitative:algorithm:pcorrectness:lemma}
If Algorithm~\ref{alg:approx_quant_reach} terminates at depth  $j$  then
\[
\yes^j(\mchain,\initstate)\leq
\probms{\mchain}(\initstate\models\eventually\fstates)\leq
\yes^j(\mchain,\initstate)+\error
\]
\end{lem}

\begin{proof}
It is straightforward to check that for each $j\geq 0$ we have
\[
\yes^j(\mchain,\initstate)\leq\probms{\mchain}(\initstate\models\eventually\fstates)
\]
and
\[
\no^j(\mchain,\initstate)\leq\probms{\mchain}(\initstate\models\unreachable{\fstates}\before\fstates)
\]
We notice that 
\[
\probms{\mchain}(\initstate\models\eventually\fstates)\leq
1-\probms{\mchain}(\initstate\models\unreachable{\fstates}\before\fstates)
\]
It follows that
\[
\yes^j(\mchain,\initstate)\leq
\probms{\mchain}(\initstate\models\eventually\fstates)
\leq
1-\no^j(\mchain,\initstate)
\]
The result follows from the fact that 
$\yes^j(\mchain,\initstate)+\no^j(\mchain,\initstate) \geq 1-\error$
when the algorithm terminates.
\end{proof}

\begin{lem}
\label{quantitative:algorithm:termination:lemma}
Algorithm~\ref{alg:approx_quant_reach} 
terminates in case the Markov chain $\mchain$ is decisive w.r.t. $\fstates$.
\end{lem}
\begin{proof}
Since $\mchain$ is decisive we have $\probms{\mchain}(\initstate\models
\eventually\fstates\vee\eventually\unreachable{\fstates})=1$.
Therefore $\lim_{j\rightarrow\infty}(\yes^j+\no^j)=1$,
which implies termination of the algorithm.
\end{proof}

From Lemma~\ref{quantitative:algorithm:pcorrectness:lemma} 
and
Lemma~\ref{quantitative:algorithm:termination:lemma} 
it follows that 
\aqr  is solvable for Markov chains which are
globally coarse w.r.t. the target set and for Markov chains which 
contain a finite attractor.
This, together with Theorem~\ref{vass:theorem} and 
Theorem~\ref{plcs:theorem}, yield the following theorems.
\begin{thm}
\label{quantitative:vass:theorem}
\aqr is solvable for {\it PVASS} in case $\fstates$ is upward-closed.
\end{thm}

\begin{thm}
\label{quantitative:plcs:theorem}
\aqr is solvable for PLCS in case $\fstates$ is effectively representable.
\end{thm}

\begin{thm}
\label{quantitative:ntm:theorem}
\aqr is solvable for PNTM in case $\fstates$ is a set of $\qntmstates$-states.
\end{thm}

\section{Approximate Quantitative Repeated Reachability}
\label{sec:quant_rep_reach}

\noindent
In this section we approximate the probability of reaching a given
set of states infinitely often, i.e., we compute arbitrarily close 
approximations of $\probms{\mchain}
\left(\initstate\models\always\eventually\fstates\right)$.

\problem{\aqrr}{
\item A Markov chain $\mchain=\mctuple$
\item A state $\initstate\in\states$
\item A set $\fstates\subseteq\states$
\item A rational $\error >0$}{
Compute a rational $\probapprox$ such that
$\probapprox\leq\probms{\mchain}
\left(\initstate\models\always\eventually\fstates\right)\leq \probapprox+\error$.}

We present an algorithm which is a modification of Algorithm~\ref{alg:approx_quant_reach}
(in Section~\ref{sec:quant_reach}) and show that it is  guaranteed to terminate 
for all Markov chains which are decisive w.r.t. both $\fstates$ and $\unreachable{\fstates}$.

\bigskip
\algbox{\aqrr\label{alg:approx_quant_repeat_reach}}
{
{\bf Input}

A Markov chain $\mchain=\mctuple$, a state $\initstate\in\states$,
a set $\fstates\subseteq\states$ and a positive $\error\in\rat_{>0}$ such that 
$\mchain$ is effective w.r.t.\ $\fstates$ and $\unreachable{\fstates}$.

{\bf Return value}

A rational $\probapprox$ such that
$\probapprox\leq\probms{\mchain}
\left(\initstate\models\always\eventually\fstates\right)\leq \probapprox+\error$

{\bf Variables}

\ind%
$\yes$, $\no$: $\rat$
\hspace{3cm}(initially all are set to 0)

\ind%
${\it store}$: queue with elements in $\states\times\rat$}{
\algline{1.}%
${\it store}:= \tuple{\initstate,1}$

\algline{2.}%
{\bf repeat}

\algline{3.}\ind%
remove $\tuple{\state,r}$ from ${\it store}$

\algline{4.}\ind%
{\bf if} $\state\in\unreachable{\unreachable{\fstates}}$ {\bf then} $\yes:=\yes+r$

\algline{5.}\ind%
{\bf else if} $\state\in\unreachable{\fstates}$ {\bf then} $\no:=\no+r$

\algline{6.}\ind%
{\bf else} \hspace{1mm}  {\bf for each} $\state'\in\post(\state)$

\algline{7.}\ind\ind%
add $\tuple{\state',r\cdot\probability(\state,\state')}$ 
to the end of ${\it store}$

\algline{8.}%
{\bf until}
$\yes+\no\geq 1-\error$

\algline{9.}%
{\bf return}
$\yes$
}

\smallskip
We require that the Markov chain is effective w.r.t.\ $\fstates$
and $\unreachable{\fstates}$
so that the conditions $\state\in\unreachable{\fstates}$ (in line 
5.) and $\state\in\unreachable{\unreachable{\fstates}}$ (in line 
4.) can be effectively checked.

We define $\yes^j(\mchain,\initstate)$ and
$\no^j(\mchain,\initstate)$ as the values of the variables $\yes$ and $\no$
after the algorithm has explored the reachability-tree 
with root $\initstate$ up-to depth $j$, similarly 
as for Algorithm~\ref{alg:approx_quant_reach}.
The following Lemma shows the partial correctness of Algorithm~\ref{alg:approx_quant_repeat_reach}.

\begin{lem}
\label{quantitative:repeated:algorithm:pcorrectness:lemma}
For a Markov chain $\mchain$ and a set $\fstates$ such that $\mchain$ 
is strongly decisive w.r.t.\ $\fstates$, if
Algorithm~\ref{alg:approx_quant_repeat_reach}
terminates at depth $j$ then
\begin{center}
$\yes^j(\mchain,\initstate)\leq
\probms{\mchain}\left(\initstate\models\always\eventually\fstates\right)
\leq\yes^j(\mchain,\initstate)+\error$
\end{center}
\end{lem}
\begin{proof}
If Algorithm~\ref{alg:approx_quant_repeat_reach} reaches some state
$\state \in \unreachable{\unreachable{\fstates}}$ (at line 4.) then we have
$\state \models \forall\always \exists\eventually \fstates$.
Since $\mchain$ is strongly decisive, it follows from
Lemma~\ref{lem:always_again_reverse} that 
$\probms{\mchain}\left(\state\models\always\eventually\fstates\right) = 1$.
Thus, for each $j\geq 0$, we have
$
\yes^j(\mchain,\initstate)\leq\probms{\mchain}
\left(\initstate\models\always\eventually\fstates\right)
$.
Similarly, if the algorithm reaches some state 
$\state \in \unreachable{\fstates}$ (at line 5.) then we have
$(s \models \always\eventually\fstates)=\emptyset$. 
Thus, for each $j\geq 0$, we have
$\no^j(\mchain,\initstate)\leq\probms{\mchain}
\left(\initstate\not\models\always\eventually\fstates\right)
=
1-\probms{\mchain}\left(\initstate\models\always\eventually\fstates\right)
$.
It follows that
$
\yes^j(\mchain,\initstate)\leq
\probms{\mchain}\left(\initstate\models\always\eventually\fstates\right)
\leq
1-\no^j(\mchain,\initstate)
$.
The result follows from the fact that 
$\yes^j(\mchain,\initstate)+\no^j(\mchain,\initstate) \geq 1-\error$
when the algorithm terminates. 
\end{proof}

\begin{lem}
\label{quantitative:repeated:algorithm:termination:lemma}
Algorithm~\ref{alg:approx_quant_repeat_reach} 
terminates if $\mchain$ is decisive w.r.t. $\unreachable{\fstates}$.
\end{lem}
\begin{proof}
Since $\mchain$ is decisive w.r.t. $\unreachable{\fstates}$, we have
$\probms{\mchain}\left(\initstate\models\eventually\unreachable{\fstates}
\vee\eventually\unreachable{\unreachable{\fstates}}\right)=1$.
It follows that $\lim_{j\rightarrow\infty}(\yes^j+\no^j)=1$, which implies
termination.
\ignore{
Let $\attractor$ be  the finite attractor. We define
$R_{\reachable{\fstates}}^{\initstate}:= \{\pi\,|\, \pi\in R_{\neg \unreachable{\fstates}}^{\initstate}\,\wedge\,\exists i.\, \pi(i) \in \reachable{\fstates}\}$ and $R_{\neg \reachable{\fstates}}^{\initstate}:= R_{\neg \unreachable{\fstates}}^{\initstate}-R_{\reachable{\fstates}}^{\initstate}$. Below, we show that 
$\probms{\mchain}(R_{\neg \reachable{\fstates}}^{\initstate})=0$
which implies the result.

All states $\state'$ visited by runs in $R_{\neg\reachable{\fstates}}^{\initstate}$ 
satisfy $\state' \models \exists\eventually\unreachable{\fstates}$.
In particular this holds for the finitely many $\state' \in A$ which are
visited by runs in $R_{\neg\reachable{\fstates}}^{\initstate}$.
Let $A' := \{\state' \in A\,|\, \exists \pi \in R_{\neg\reachable{\fstates}}^{\initstate}\,\exists
i.\, \pi(i)=\state'\}$.
For every $\state' \in A'$ we define 
$\alpha_{\state'} := \probms{\mchain}\left(\state' \models
\eventually\unreachable{\fstates}\right)$ and obtain $\alpha_{\state'} >0$.
By finiteness of $A$ (and $A'$) we get $\alpha := {\it min}_{\state' \in A'}
\alpha_{\state'} > 0$. As $A$ is a finite attractor, almost every run
in $R_{\neg\reachable{\fstates}}^{\initstate}$ must visit
$A$ (and thus $A'$) infinitely often. 
Since $R_{\neg\reachable{\fstates}}^{\initstate}\subseteq R_{\neg\unreachable{\fstates}}^{\initstate}$,
 it follows that the set of runs  in $R_{\neg \reachable{\fstates}}^{\initstate}$ has measure 
$\le (1-\alpha)^\infty = 0$.} 
\end{proof}

Note that Algorithm~\ref{alg:approx_quant_repeat_reach} only works for Markov
chains which are decisive w.r.t.\ both $\fstates$ and $\unreachable{\fstates}$.
Decisiveness w.r.t.\ $\unreachable{\fstates}$ is required for termination 
(Lemma~\ref{quantitative:repeated:algorithm:termination:lemma}),
while decisiveness w.r.t.\ $\fstates$ is required for correctness
(Lemma~\ref{quantitative:repeated:algorithm:pcorrectness:lemma}).

Now we show the computability results for PLCS and PNTM.

\begin{thm}
\label{quantitative:repeated:plcs:theorem}
\aqrr is solvable for PLCS in case $\fstates$ is effectively representable.
\end{thm}
\begin{proof}
By Theorem~\ref{plcs:theorem}, a Markov chain induced by a PLCS is decisive 
w.r.t. every set, in particular w.r.t. $\fstates$ and
$\unreachable{\fstates}$.
Thus it follows from Lemma~\ref{quantitative:repeated:algorithm:pcorrectness:lemma} and
Lemma~\ref{quantitative:repeated:algorithm:termination:lemma},
that Algorithm~\ref{alg:approx_quant_repeat_reach} solves \aqrr for PLCS.
\end{proof}

\begin{thm}
\label{quantitative:repeated:ntm:theorem}
\aqrr is solvable for PNTM in case $\fstates$ is a set of $\qntmstates$-states.
\end{thm}
\begin{proof}
Assume a PNTM $\ntm=\ntmtuple$ and a set $\fstates$ of $\qntmstates$-states for some 
$\qntmstates\subseteq\ntmstates$.
Since $\fstates$ is the set of $\qntmstates$-states, it follows by Lemma~\ref{lem:ntm_ctl} 
that $\unreachable{\fstates}$ is also a set of $\qntmstates'$-states where 
$\qntmstates'\subseteq\ntmstates$.
By Theorem~\ref{ntm:theorem}, the Markov chain induced by a PNTM is decisive 
w.r.t. every set of $\qntmstates''$-states, in particular w.r.t. $\fstates$ and
$\unreachable{\fstates}$.
Thus it follows from Lemma~\ref{quantitative:repeated:algorithm:pcorrectness:lemma} and
Lemma~\ref{quantitative:repeated:algorithm:termination:lemma},
that Algorithm~\ref{alg:approx_quant_repeat_reach} solves \aqrr for PNTM.
\end{proof}

A similar result for PVASS would require the explicit assumption that the
induced Markov chain is decisive w.r.t.\ $\fstates$ and
$\unreachable{\fstates}$. It is not sufficient that $\fstates$ is
upward-closed, because this only implies decisiveness w.r.t.\ $\fstates$,
not necessarily w.r.t.\ $\unreachable{\fstates}$. (See the counterexample
in Remark~\ref{rem:dec_F_Ftilde}.)

\section{Exact Quantitative Analysis}
\label{sec:exact_quant_reach}

\noindent
In this section we consider the {\it Exact
Quantitative Reachability Analysis Problem}, defined as follows.

\ignore{
\vspace{2mm}
\noindent
\framebox{
\begin{minipage}{0.90\hsize}
\eqr
\\
{\bf Instance}
A Markov chain $\mchain=\mctuple$, a state $\initstate\in\states$, 
a set of states $\fstates\subseteq\states$
and a rational $\probbound$.
\\
{\bf Task}
Check whether
$\probms{\mchain}
\left(\initstate\models\eventually\fstates\right)\geq\probbound$.
\end{minipage}
}
\vspace{1mm}
}

\problem{\eqr}{
\item A Markov chain $\mchain=\mctuple$
\item A state $\initstate\in\states$
\item A set of states $\fstates\subseteq\states$
\item A rational $\probbound$}{
Check whether
$\probms{\mchain}
\left(\initstate\models\eventually\fstates\right)\geq \probbound$.}

By Theorem~\ref{thm:undecidable_upwardclosed}, \eqr is undecidable for
PVASS and upward-closed sets $\fstates$. If $\fstates$ is a set of
$\qvassstates$-states then decidability of 
\eqr is open for PVASS, PLCS and PNTM.

However, for PVASS, PLCS and PNTM, we show that the probability
$\probms{\mchain}\left(\initstate\models\eventually\fstates\right)$
(and thus the question of \eqr)
cannot be effectively expressed in a uniform way in the first-order theory of the reals
$(\R,+,*,\le)$, or any other decidable logical theory with first-order
quantification. By `expressed in a uniform way' we mean
that parameters of the system (e.g., transition weights for PVASS, the message
loss probability for PLCS or the noise parameter for PNTM) should be 
reflected directly by constants in the
constructed formula (see the remarks at the end of this section for details).
This negative result for PVASS, PLCS and PNTM is in contrast to the situation for 
probabilistic pushdown automata (probabilistic recursive systems)
for which this probability can be effectively
expressed in $(\R,+,*,\le)$ \cite{EsparzaKuceraMayr:lics2004,EKM:LMCS2006,Etessami-Yannakakis:STACS05,Esparza:Etessami:fsttcs04K}.

\begin{thm}\label{thm:not_constructible}
Let $\vass=\pvasstuple$ be a PVASS and $\vassweight_1,\dots,\vassweight_n \in \rat$ the
constants used in the transition weight function $\vassweight$. Let 
$\fstates$ be the set of $\qvassstates$-states for some
$\qvassstates \subseteq \vassstates$. 

It is impossible to effectively
construct a {$(\R,+,*,\le)$} formula $\Phi(p,\vassweight_1,\dots,\vassweight_n)$ with parameters
$\vassweight_i$ and $p$ which expresses the probability 
$\probms{\mchain}\left(\initstate\models\eventually\fstates\right)$, i.e.,
$\Phi(p,\vassweight_1,\dots,\vassweight_n) = {\it true}$ iff 
$p = \probms{\mchain}\left(\initstate\models\eventually\fstates\right)$. 
\end{thm}
\begin{proof}
We assume the contrary and derive a contradiction. 
Consider the IT-2-CM $M$ and the PVASS $\vass$ with parameter $x$ 
from Def.~\ref{def:2cm_pvass}, i.e., let $\vassweight_1=x$ and $\vassweight_i=1$ for $i > 1$.
Let $\fstates$ be the 
set of ${\it err}$-states. Suppose that one could effectively construct the
$(\R,+,*,\le)$-formula $\Phi(p,\vassweight_1,\dots,\vassweight_n)=\Phi(p,x)$ with the required
properties.

If a counter is tested for zero then, in our weak simulation by the PVASS,
there are two cases:
\begin{enumerate}[$\bullet$]
\item
If the counter contains zero, then only one transition (the one of type 
$\beta$) is enabled and the simulation is faithful.
After firing transition $\beta$, only transition $\gamma$ (but not $\delta$)
is enabled.
\item
If the counter does not contain zero then two transitions, $\alpha$ and $\beta$,
are enabled. The probability of choosing $\alpha$, the faithful simulation,
is $1/(1+x)$ and the probability of choosing $\beta$ (the wrong transition in
this case) is $x/(1+x)$.
If $\beta$ is fired then both $\gamma$ and $\delta$ are enabled.
\end{enumerate}

If the IT-2-counter machine $M$ accepts after a finite number $L$ of steps,
then it can make at most $L$ tests for zero. Thus the probability of ever
choosing the wrong transition is bounded from above by 
$1-1/(1+x)^L$. A transition of type $\delta$, leading 
to the ${\it err}$-state can only be taken if a wrong transition has been taken first.
Thus the probability of reaching the ${\it err}$-state is also bounded 
from above by
$1-1/(1+x)^L$. For $x \rightarrow 0$ this probability converges to $0$.
Thus we have $\exists x>0\,\exists p\,(\Phi(p,x) \wedge p < 1/10)$.

If the IT-2-counter machine $M$ does not accept then its infinite run will
contain infinitely many tests for zero on counters which are non-zero 
by Def.~\ref{def:2cm_pvass}.
In each of those tests, the chance of firing the wrong transition 
(type $\beta$) is $x/(1+x) > 0$. Thus it will happen eventually with probability 1. 
If the wrong transition has been fired then the probability of going to the
${\it err}$-state by the next transition is $1/2$ (competing enabled
transitions $\gamma$ and $\delta$ of weight 1 each).
Thus the probability of reaching the ${\it err}$-state is at least $1/2$, i.e.,
$\probms{\mchain}\left(\initstate\models\eventually\fstates\right) \ge 1/2$, 
regardless of the value of $x$ (provided $x>0$).

It follows that $M$ accepts if and only if 
\[
\exists x>0\,\exists p\,(\Phi(p,x) \wedge p < 1/10)
\]
If the formula $\Phi(p,x)$ was effectively constructible, then 
the formula $\exists x>0\,\exists p\,(\Phi(p,x) \wedge p < 1/10)$ would
also be effectively constructible and the question 
would be decidable, because of the decidability of $(\R,+,*,\le)$ 
\cite{Tarski:reals-arithmetic}. 
This is a contradiction, since acceptance of $M$ is undecidable.
\end{proof}

\begin{rem}
In Theorem~\ref{thm:not_constructible} the set $\fstates$ is the set of 
${\it err}$-states, i.e., an upward- and downward-closed set. However, the 
result holds just as well if $\fstates$ is a single configuration.
To show this, it suffices to modify the construction as follows.
Add two new transitions $({\it err}, \op_i, {\it err})$ with weight 1 and
$\op_i(c_i)=-1$ and $\op_i(c_j)=0$ for $i\neq j$ and $i,j \in \{1,2\}$.
The only other possible transition in ${\it err}$-states is the
self-loop which changes nothing. So almost every run starting in an
${\it err}$-state will eventually reach $({\it err},0,0)$.
Thus the probability of eventually reaching configuration $({\it err},0,0)$ 
is the same as that of eventually reaching some ${\it err}$-state.
\end{rem}

For PLCS, the result corresponding to Theorem~\ref{thm:not_constructible}
would be trivial if one allowed the case that the message loss probability
$\lambda$ is zero, because the reachability problem for deterministic non-lossy 
non-probabilistic FIFO-channel systems is undecidable \cite{Brand:CFSM}
(unlike for VASS \cite{Mayr:SIAM84}).
The following theorem shows a stronger non-expressibility result even for
the restricted case of $\lambda >0$. 

\begin{thm}\label{thm:plcs_not_constructible}
Let $\plcs=\plcstuple$ be a PLCS with message loss probability $\lambda >0$ and 
$\fstates$ a set of $\qvassstates$-states for some
$\qvassstates \subseteq \vassstates$. 
Then it is impossible to effectively
construct a {$(\R,+,*,\le)$} formula $\Phi(p,\lambda)$ with parameters
$\lambda>0$ and $p$ which expresses the probability 
$\probms{\mchain}\left(\initstate\models\eventually\fstates\right)$, i.e.,
such that for any $\lambda >0$ one has $\Phi(p,\lambda) = {\it true}$ iff 
$p = \probms{\mchain}\left(\initstate\models\eventually\fstates\right)$. 
\end{thm}
\begin{proof}
We assume the contrary and derive a contradiction. It is known that 
the termination problem for deterministic non-lossy 
non-probabilistic FIFO-channel systems $\plcs'$ is undecidable \cite{Brand:CFSM}.
Given such a FIFO-channel system $\plcs'$ (and the contrary of our theorem)
we will construct a PLCS $\plcs$ and a {$(\R,+,*,\le)$} formula $\Psi$
s.t. $\Psi$ is true if and only if $\plcs'$ terminates.

Consider a deterministic non-lossy non-probabilistic FIFO-channel system
$\plcs'$ which starts in control-state $\initstate$ and the empty channel.
Let $k_{\it acc}$ be the final accepting control-state of $\plcs'$.
$\plcs'$ either eventually reaches the final accepting control-state $k_{\it acc}$ and terminates,
or continues forever. We construct the PLCS $\plcs$ by modifying $\plcs'$ as follows.
First we add a new control-state ${\it err}$ to the system. Then, for
every transition $t_1$ of $\plcs'$ we add an additional transition $t_2$ to
$\plcs$. Transition $t_2$ is almost identical to $t_1$, except that its target
control-state is ${\it err}$. We assign the same transition weight 1 to every
transition in $\plcs$. Thus, in every single step, the system $\plcs$ has 
probability $1/2$ of going to an ${\it err}$-state.
In order to avoid deadlocks, 
we add self-loops to the control-states $k_{\it acc}$ and ${\it err}$.
These are the only possible transitions from $k_{\it acc}$ and ${\it err}$.
In particular, it is impossible to get to an ${\it err}$-state from any 
$k_{\it acc}$-state or vice-versa.
Finally, we add the message loss probability $\lambda >0$ to $\plcs$, i.e.,
in every step any message in transit is lost with probability $\lambda$.
Let $\fstates$ be the set of $k_{\it acc}$-states.

If $\plcs'$ terminates, then it terminates in a finite number $L$ of steps.
Since one starts with the empty channel, the maximal number of messages
in transit at any time in this run is bounded from above by $L$.
The PLCS $\plcs$ can imitate this run of $\plcs'$ (however, $\plcs$ 
also has many other possible runs).
The probability that none of the (at most $L$) messages is lost in any single
step is bounded from below by $(1-\lambda)^L$.
Thus $\probms{\mchain}\left(\initstate\models\eventually\fstates\right)
\ge \left((1-\lambda)^L \cdot \frac{1}{2}\right)^L$. This is the probability
that $\plcs$ faithfully simulates the behavior of $\plcs'$. No messages are lost and
all transitions leading to ${\it err}$-states are avoided.
Now let $\epsilon := (0.9)^{L^2} (0.5)^L$. It follows that
\[
\exists\epsilon >0.\, \forall \lambda : 0 < \lambda \le 0.1\, 
\probms{\mchain}\left(\initstate\models\eventually\fstates\right) \ge \epsilon
\]
Now we consider the case that $\plcs'$ does not terminate. Let $G$ be the set
of all ${\it err}$-states. In particular, $G \subseteq \unreachable{\fstates}$.
Since $G$ is not reachable from $\fstates$ either, we have 
$\probms{\mchain}\left(\initstate\models\neg\eventually\fstates\right) \ge
\probms{\mchain}\left(\initstate\models\eventually G\right)$.
Consider all those runs of $\plcs$ which do not lose any messages
in the first $N$ steps
and reach $G$ after at most $N$ steps. These runs faithfully
simulate the first $N$ steps of $\plcs'$, unless they go 
to an ${\it err}$-state. In particular, they do not reach $k_{\it acc}$,
because $\plcs'$ does not reach $k_{\it acc}$, since $\plcs'$ does not
terminate.
We have
\[
\probms{\mchain}\left(\initstate\models\eventually G\right) \ge
(1-\lambda)^{N^2} \left(1-(0.5)^N\right)
\]
Now let
$N:=\left(\frac{1}{\lambda}\right)^{1/4}$.
We obtain 
\[
\lim_{\lambda \rightarrow 0}
\probms{\mchain}\left(\initstate\models\eventually G\right) = 1
\]
and thus 
\[
\lim_{\lambda \rightarrow 0}
\probms{\mchain}\left(\initstate\models\eventually \fstates\right) = 0
\]
and finally
\[
\neg\left(\exists\epsilon >0.\, \forall \lambda : 0 < \lambda \le 0.1\, 
\probms{\mchain}\left(\initstate\models\eventually\fstates\right) \ge \epsilon\right)
\]
Therefore, $\plcs'$ terminates if and only if
\[
\exists\epsilon >0.\, \forall \lambda : 0 < \lambda \le 0.1\, 
\probms{\mchain}\left(\initstate\models\eventually\fstates\right) \ge \epsilon
\]
Assume that for $\plcs$ the {$(\R,+,*,\le)$}-formula $\Phi(p,\lambda)$ was 
effectively constructible. Then $\plcs'$ terminates if and only if
\[
\Psi := \exists\epsilon >0.\, \forall \lambda : 0 < \lambda 
\le 0.1\, \exists p\, (\Phi(p,\lambda) \wedge p \ge \epsilon) 
\]
Since $(\R,+,*,\le)$ is decidable \cite{Tarski:reals-arithmetic} one can
decide if $\Psi$ is true and thus if $\plcs'$ terminates. Contradiction.
\end{proof}

The proof of the corresponding result for PNTM is similar to the one for PLCS.
Note that the following Theorem~\ref{thm:ntm_not_constructible} also holds for
NTM, since the proof does not use the extensions of PNTM over NTM.

\begin{thm}\label{thm:ntm_not_constructible}
Let $\ntm=\ntmtuple$ be a (P)NTM with noise probability $\noise >0$ and 
$\fstates$ a set of $\qntmstates$-states for some
$\qntmstates \subseteq \ntmstates$. 
Then it is impossible to effectively
construct a {$(\R,+,*,\le)$} formula $\Phi(p,\noise)$ with parameters
$\noise>0$ and $p$ which expresses the probability 
$\probms{\mchain}\left(\initstate\models\eventually\fstates\right)$, i.e.,
such that for any $\noise >0$ one has $\Phi(p,\noise) = {\it true}$ iff 
$p = \probms{\mchain}\left(\initstate\models\eventually\fstates\right)$. 
\end{thm}
\begin{proof}
We assume the contrary and derive a contradiction.
In the proof, we start form a Turing machine $\ntm'$. 
Then we construct an PNTM $\ntm$ and a {$(\R,+,*,\le)$}-formula $\Psi$
s.t. $\Psi$ is true if and only if $\ntm'$ halts.
The contradiction follows from the undecidability of the halting 
problem for Turing-machines~\cite{HoUl:book}.

Assume a single-tape Deterministic Turing machine $\ntm'$ that starts in an initial control-state 
$\ntmstate_0$ with an input word $\omega_0$. 
Once started, the machine either reaches a control-state 
$\ntmstate_h$ and halts, or runs forever.
We derive a PNTM $\ntm$ from $\ntm'$ as follows.
First, we add a control-state $\ntmstate_e$ to the system and a special symbol 
$\ntmsymbol$ to the tape alphabet.
Then, from every state, we add a transition that reads $\ntmsymbol$ and moves to 
$\ntmstate_e$ where it halts.
We add loops to $\ntmstate_e$ and $\ntmstate_h$ to avoid deadlocks.
Finally, we assume that $\ntm$ has a noise probability $\noise>0$ and all the 
transitions have weight $1$.

Assume that the runs of $\ntm$ start all from an initial (global) state 
$\initstate$ where the control-state is $\ntmstate_0$ and the initial word is
$\omega_0$.
Let $\fstates$ be the set of $\ntmstate_h$-states. 
If $\ntm'$ halts then it halts after a finite number of steps $L$.
The runs which faithfully imitate those of $\ntm'$ avoid noise in each step.
Assume that the computation reached a step where the cell under the head was 
visited $k$ time units ago.
Then, the probability to avoid the noise and move to the next configuration 
is exactly $1-\sum_{\ntmsymbol'\in\tapealphabet}\frac{1-(1-\noise)^k}{|\tapealphabet|}=(1-\noise)^k$. 
Observe that this probability is bounded from below by $(1-\noise)^L$.
Therefore, it follows that $\probms{\mchain}\left(\initstate\models\eventually\fstates\right)\ge(1-\noise)^{L^2}$.
Thus, by taking $\theta=0.9^{L^2}$ we obtain
\[
\exists\theta >0.\, \forall \noise : 0 < \noise \le 0.1\, 
\probms{\mchain}\left(\initstate\models\eventually\fstates\right) \ge \theta.
\]

Suppose that $\ntm'$ does not halt and continues running forever. 
Let $G$ be the set of $\ntmstate_e$-states. 
Observe that $\fstates$ is not reachable from $G$.
Similarly, by reasoning about runs of $\ntm$ which faithfully imitate those of 
$\ntm'$ the first $N$ steps, we obtain that 
$\probms{\mchain}\left(\initstate\models\neg\eventually\fstates\right)\ge\probms{\mchain}\left(\initstate\models\eventually G\right)\ge(1-\noise)^{N^2}$.
For $N=(\frac{1}{\noise})^{1/4}$, it follows that $\lim_{\noise\rightarrow 0}\probms{\mchain}\left(\initstate\models\eventually G\right)=1$.
As a consequence, we obtain $\lim_{\noise\rightarrow 0}\probms{\mchain}\left(\initstate\models\eventually\fstates\right)=0$.
Therefore, we have 
\[
\neg\left(\exists\theta >0.\, \forall \noise : 0 < \noise \le 0.1\, 
\probms{\mchain}\left(\initstate\models\eventually\fstates\right) \ge \theta\right).
\]

Finally we obtain that $\ntm'$ halts if and only if 
\[
\exists\theta >0.\, \forall \noise : 0 < \noise \le 0.1\, 
\probms{\mchain}\left(\initstate\models\eventually\fstates\right) \ge \theta.
\]
Suppose that for $\ntm$, there exists a {$(\R,+,*,\le)$}-formula $\Phi(p,\noise)$.
Then $\ntm'$ terminates if and only if
\[
\Psi:=\exists\theta >0.\, \forall \noise : 0 < \noise \le 0.1\, 
\exists p\,(\Phi(p,\noise)\wedge p\ge \theta)
\]
which is a contradiction.
\end{proof}

\begin{rem}\label{rem:rep_reach_nonexp}
In the constructions for Theorem~\ref{thm:not_constructible},
Theorem~\ref{thm:plcs_not_constructible} 
and Theorem~\ref{thm:ntm_not_constructible}
only states in $\fstates$ can be
reached from $\fstates$.
Hence
$
\probms{\mchain}\left(\initstate\models\eventually\fstates\right)
=
\probms{\mchain}\left(\initstate\models\always\eventually\fstates\right)
$
in these constructions. 
Thus, neither for PVASS, nor for PLCS or PNTM the probability
$\probms{\mchain}\left(\initstate\models\always\eventually\fstates\right)$ can
be expressed effectively in $(\R,+,*,\le)$.
\end{rem}

To summarize, the results of this section only show that for PVASS, PLCS
and (P)NTM the probabilities
$\probms{\mchain}\left(\initstate\models\eventually\fstates\right)$
and 
$\probms{\mchain}\left(\initstate\models\always\eventually\fstates\right)$
cannot be effectively expressed in $(\R,+,*,\le)$ in the same uniform way as
for probabilistic pushdown automata (see 
\cite{EsparzaKuceraMayr:lics2004,EKM:LMCS2006,Etessami-Yannakakis:STACS05,Esparza:Etessami:fsttcs04K}).

The following three scenarios are not ruled out for PVASS, PLCS, or PNTM.\/ 
It is an open question which of them is true for each class of models.
\begin{enumerate}[(1)]
\item
The probability
$\probms{\mchain}\left(\initstate\models\eventually\fstates\right)$
is effectively expressible in $(\R,+,*,\le)$, but not in a uniform way.
For example, the parameters of the (PVASS, PLCS or PNTM) system would influence the formula in a 
complex (but still computable) way such that it is not possible to
quantify over them in the logic.
\item
The probability
$\probms{\mchain}\left(\initstate\models\eventually\fstates\right)$
is expressible in $(\R,+,*,\le)$, but the formula is not effectively
constructible.
\item
The probability
$\probms{\mchain}\left(\initstate\models\eventually\fstates\right)$
is not generally expressible in $(\R,+,*,\le)$, i.e., there is
no corresponding formula at all.
\end{enumerate}
  
\section{Conclusions and Future Work}
\label{sect:conclusions}

\noindent
We have defined the decisiveness property for infinite-state 
Markov chains. Some other abstract conditions like the existence of
a finite attractor or global coarseness imply the decisiveness property.
In particular, several classes of infinite Markov chains which are derived
from program-like probabilistic system models like PVASS, PLCS and PNTM 
satisfy these properties.
 
We have studied the qualitative and quantitative (repeated) reachability
problem for decisive Markov chains. Our main results are that
qualitative questions (i.e., if some probability is $0$ or $1$)
can often be reduced to questions about the underlying infinite transition
graph of the systems, while for quantitative questions a simple path 
enumeration algorithm can be used to approximate the probabilities 
arbitrarily closely.
 
A surprising result was that reachability of control-states
and reachability of upward-closed sets cannot be effectively expressed in
terms of each other for PVASS, unlike for normal VASS
(Section~\ref{sec:quali_reach}). Furthermore, for probabilistic systems, 
reachability is not always easier to decide than repeated reachability
(Theorems~\ref{thm:undecidable_upwardclosed} and
\ref{thm:0_1_upwardclosed_repeat}). 

Open questions for future work are the decidability of qualitative
reachability problems for Markov chains with downward-closed sets of final states, 
and an algorithm to approximate quantitative repeated reachability 
in PVASS.\/ Furthermore, the decidability of exact quantitative questions
like $\probms{\mchain}\left(\initstate\models\eventually\fstates\right) \ge
0.5$ is still open for PVASS, PLCS and (P)NTM.

\bibliographystyle{alpha}
\bibliography{ref}

\end{document}